\documentclass[english,aps,prx,twocolumn,reprint]{revtex4-2}
\usepackage[T1]{fontenc}
\usepackage{textcomp}
\usepackage[latin9]{inputenc}
\usepackage{color}
\usepackage{babel}
\usepackage{units}
\usepackage{amsmath}
\usepackage{amssymb}
\usepackage{graphicx}
\usepackage[bookmarks=false,
 breaklinks=false,pdfborder={0 0 1},backref=false,colorlinks=false]
 {hyperref}

\makeatletter

\providecommand{\tabularnewline}{\\}

\usepackage{physics}

\makeatother

\begin{document}
\title{Generalized Linear Response Theory for Pumped Systems\\
and its Application to Transient Optical Properties}
\author{Amir Eskandari-asl$^{1}$ and Adolfo Avella$^{1,2,3}$}
\address{$^{1}$Dipartimento di Fisica ``E.R. Caianiello'', Università degli
Studi di Salerno, I-84084 Fisciano (SA), Italy}
\address{$^{2}$CNR-SPIN, Unità di Salerno, I-84084 Fisciano (SA), Italy}
\address{$^{3}$CNISM, Unità di Salerno, Università degli Studi di Salerno,
I-84084 Fisciano (SA), Italy}
\begin{abstract}
We derive the two-time linear response theory for out-of-equilibrium
pumped systems, generic pump-probe delays and probe frequencies. Such
a theory enormously simplifies the numerical calculations, for instance,
of the optical conductivity with respect to the actual procedure,
which requires computing the effect of the probe pulse for each time
delay with respect to the pump pulse\textcolor{red}{.} The theory
is given for a generic observable and pumped Hamiltonian and then
specialized for a system with a quadratic Hamiltonian and its transient
optical properties, exploiting the Dynamical Projective Operatorial
Approach (DPOA). The theory is complemented by a set of crucial numerical
guidelines that help perform actual calculations in a computationally
affordable way. The optical response (differential transient reflectivity
and absorption) of a prototypical three-band (core, valence, and conduction)
model in the XUV regime is analyzed in detail to illustrate the theory
and its application. Using some generalizations of the density of
states, we provide a systematic approach to exploring the optical
properties in terms of the system band structure features and the
pump parameters. Such an analysis can be extremely helpful in understanding
the actual results of experimental optical measurements. Moreover,
we study the effects of inter-band and intra-band transitions, the
local dipole coupling, and single and multi-photon processes. The
latter is further investigated by varying the central frequency of
the pump pulse to have different regions of the first Brillouin zone
in resonance with it. We also study the effect of varying the pump
pulse intensity. Finally, we study and analyze the transient optical
properties in the probe pulse regime of IR and visible.
\end{abstract}
\maketitle

\section{Introduction}

Thanks to modern technological developments, studying the real-time
dynamics of condensed matter systems in the ultra-fast attosecond
regime is now possible. For such studies, one deals with the so-called
pump-probe setups, where a system is driven out-of-equilibrium by
an intense ultra-short electromagnetic pump pulse and is investigated
using a rather low-intensity probe pulse, as a function of the time
delay between the two pulses \citep{brabec2000intense,krausz2009attosecond,krausz2014attosecond,calegari2016advances,borrego2022attosecond}.
The study of the real-time dynamics of ultra-fast electronic excitation
in the attosecond regime unravels the fundamental physical transitions
in the pumped system \citep{Zurch_17,doi:10.1063/1.4985056,PhysRevB.97.205202,perfetti2008femtosecond,schmitt2008transient}.
This knowledge is relevant for fundamental physics and future ultra-fast
electro-optical circuit and device engineering.

Several different approaches can experimentally investigate the pumped
systems. One way is to study the high harmonic generation (HHG), where
one studies the radiations generated by the pumped system with frequencies
of integer multiples of the pump frequency \citep{von1995generation,norreys1996efficient,chin2001extreme,ghimire2011observation,luu2015extreme,han2016high,Borja:16,you2017anisotropic,liu2017high,jiang2019crystal,schmid2021tunable,neufeld2022probing,Heide:22,jordi2022hhs,Li2023reson,zhang2024high}.
Another approach to studying the pumped system is the time-resolved
angle-resolved photoemission spectroscopy (TR-ARPES) \citep{schmitt2008transient,rohwer2011collapse,smallwood2012tracking,hellmann2012time,papalazarou2012coherent,wang2013observation,johannsen2013direct,rameau2016energy,reimann2018subcycle,Soifer2019band,schuler2021theory,Neufeld2022time,ito2023build,Neufeld_2024},
where analyzing the energy and momentum distribution of the electrons
ejected from the pumped system via photoelectric effect as a function
of the time delay between the pump and probe pulses, one determines
the out-of-equilibrium electronic properties. Another way for studying
the pumped systems is through measuring the transient dynamical variations
of their optical responses to a weak probe pulse \citep{schultze2013controlling,stojchevska2014ultrafast,schultze2014attosecond,lucchini2016attosecond,mashiko2016petahertz,Borja:16,zurch2017ultrafast,Zurch_17,schlaepfer2018attosecond,Kaplan:19,geneaux2019,inzani2022field}.
In this work, we focus on the latter and provide a convenient theoretical
tool to facilitate this analysis.

For studying the transient optical properties, one usually studies
the dynamical variation of the absorption coefficient \citep{schultze2013controlling,schultze2014attosecond,lucchini2016attosecond,mashiko2016petahertz,Borja:16,zurch2017ultrafast,Zurch_17,schlaepfer2018attosecond,geneaux2019}
or the reflectivity \citep{schultze2013controlling,stojchevska2014ultrafast,Kaplan:19,inzani2022field,inzani2023photoinduced}
of the probe pulse as a function of the probe-pulse frequency (or
equivalently, probe photon energy) and the time delay between the
pump and probe pulses. Varying the photon energy of the probe pulse,
one can study different energy ranges of the band structure and their
corresponding dynamical excitation transitions. A very convenient
optical approach to reveal the pumping mechanisms in solids is to
use a probe pulse with a much higher frequency than the pump pulse.
In such a scenario, an intense low-frequency (e.g., in the IR or visible
ranges) laser pulse pumps the system out of equilibrium by altering
the electronic bands and distributions for the energies near the Fermi
level. At the same time, the strongly bounded \emph{core} electrons
are left almost unaltered. The high-frequency (e.g., in the XUV range)
probe pulse then scans those out-of-equilibrium electronic levels
by exciting the \emph{core} electrons to them \citep{schultze2013controlling,schultze2014attosecond,lucchini2016attosecond,mashiko2016petahertz,Borja:16,zurch2017ultrafast,Zurch_17,schlaepfer2018attosecond,Kaplan:19,geneaux2019,inzani2022field}.
The recent experimental setups exploit pump pulses with durations
of the order of tens of femtoseconds and probe them with even shorter
delay times \citep{schultze2013controlling,schultze2014attosecond,lucchini2016attosecond,mashiko2016petahertz,Borja:16,schlaepfer2018attosecond,geneaux2019,inzani2022field}.
Such ultra-short time intervals rule out several relaxation transitions
on longer time scales, including spontaneous emission, electron-phonon
interaction, etc. This is of fundamental interest as one can study
pure early-time electronic dynamics with several potential technological
applications, such as manufacturing high-speed electronics.

On the theory side, the standard approach to numerically simulate
the pumped systems is the time-dependent density functional theory
(TD-DFT) \citep{de2013inside,de2016monitoring,wopperer2017efficient,de2017first,pemmaraju2018velocity,tancogne2018atomic,de2018real,schlaepfer2018attosecond,inzani2022field,neufeld2023attosecond},
which is capable of considering different kinds of interactions up
to some approximations. However, there are two relevant drawbacks
in TD-DFT: (i) the computational cost is very high, and (ii) it's
more complex to get deep insight into the underlying physics of the
results of the simulations. On the other hand, the model-Hamiltonian
approaches, which generally allow a better understanding of the underlying
physical phenomena, usually suffer from the problem of oversimplification
\citep{armstrong2021dialogue,schlaepfer2018attosecond}.

To overcome some of the limitations and drawbacks of the current theoretical
approaches, we have developed a method, dubbed the Dynamical Projective
Operatorial Approach (DPOA), and applied it to a pumped germanium
sample \citep{inzani2022field}. We benchmarked our results with those
of TD-DFT on a rough momentum grid (to be affordable for TD-DFT).
Then, we extended our study to much finer momentum grid, showing that
the latter was necessary to truly understand the complicated transitions
in the actual material and clarify the individual roles of different
mechanisms and their interplay effects. Moreover, being capable of
computing all multi-time multi-particle correlation functions using
DPOA, we extended the theory to study TR-ARPES. We showed how one
can study the effects and interplays of different pumping mechanisms,
as well as the role of crystal symmetry in the TR-ARPES signal, and
how various types of sidebands emerge \citep{eskandari2023dynamical}.
Very preliminary results of the application of DPOA to pumped strongly
correlated systems can be found in Ref.~\citep{eskandari2024out}.

Theoretical study and numerical simulation for the optical properties
is much heavier than the other scenarios (HHG and TR-ARPES), as for
each value of the time delay between the pump and probe pulses, one
needs to compute the current in the presence of both pulses and then
subtract the current induced by the pump pulse only from it, to obtain
the two-time optical conductivity that can be Fourier transformed
to get the out-of-equilibrium dielectric function and, finally, the
transient reflectivity and absorption. Given that in the experimental
setups, they consider very many time delays, performing TD-DFT simulations
becomes extremely heavy and challenging, and one usually needs to
use a spare sampling of the first Brillouin zone (FBZ) by exploiting
a rough momentum grid, which may indeed result in losing some relevant
features in the simulations.

Instead, one can get a noticeable simplification by exploiting an
out-of-equilibrium version of the linear response theory (LRT), where
the two-time linear response of the pumped system to the probe pulse
is computed. This latter is the core of this work, where the out-of-equilibrium
LRT formulation is given for a general case. Then, it is specialized
for a pumped lattice system with an effective quadratic Hamiltonian
within DPOA. After that, we apply the out-of-equilibrium LRT to obtain
the optical conductivity and, consequently, the dielectric function,
from which one, in principle, can calculate the reflectivity and absorption
coefficient. We apply this theory to a prototypical three-band model
and study the different out-of-equilibrium optical behaviors that
can emerge in various regimes.

The rest of this article is organized as follow. In Sec.~\ref{sec:Theory},
we report the derivation of the two-time linear response theory for
an out-of-equilibrium pumped system (Sec.~\ref{subsec:Linear-response-in}),
its specialization to lattice systems with quadratic Hamiltonians
(Sec.~\ref{subsec:Pumped-lattice-systems}), and to their optical
conductivity (Sec.~\ref{subsec:Optical-conductivity}), reflectivity
and absorption (Sec.~\ref{subsec:Reflectivity-and-Absorption}),
and delve into the equilibrium case too (Sec.~\ref{subsec:Equilibrium-properties}).
We also report some important guidelines for numerical calculations
(Sec.~\ref{subsec:Guidelines-for-numerical}): a simplified and numerically
effective rewriting of the optical conductivity expression (Sec.~\ref{subsec:Rewriting-sigma-xyz}),
a discussion on how to choose the value of the unavoidable damping
factor (Sec.~\ref{subsec:Finite-k-grid}), a possible analytical
simplification, leading to a significative reduction of the calculation
time, regarding the time range after the application of the pump pulse
(Sec.~\ref{subsec:Analytical-after-pump}), two possible approximations
regarding the core levels (Sec.~\ref{subsec:Approximations-core-levels})
and a quasi-static dynamics (Sec.~\ref{subsec:QS-main-text}). In
Sec.~\ref{sec:Two-band_system}, we define (Sec.~\ref{subsec:The-system})
and study a prototypical three-band (core, valence, and conduction)
model in presence of only the Peierls substitution coupling (Sec.~\ref{subsec:Peierls-substitution-coupling})
and analyze the effects of inter- and intra-band transitions driven
by the pump pulse in the model (Sec.~\ref{subsec:Inter-and-intra-band}).
After that, we consider both the Peierls substitution and the dipole
couplings (Sec.~\ref{subsec:Dipole-coupling}) to study the features
originating from the latter. Then, going back to the case of having
only the Peierls substitution coupling, we study the dependence of
the observed features on the frequency (Sec.~\ref{subsec:Pump-pulse-frequency-dependence})
and the intensity (Sec.~\ref{subsec:Pump-pulse-intensity-dependence})
of the pump pulse. Moreover, we investigate the optical properties
in the probe pulse ranges of the infrared (IR) and visible (Sec.~\ref{subsec:IR-and-visible}).
Sec.~\ref{sec:Summary-and-conclusions} summarizes this work. Finally,
we have a plethora of appendices discussing: the linear variation
of the density matrix in the presence of the probe pulse (Sec.~\ref{sec:DM-linear-variation}),
the dielectric function and the reflectivity formula out of equilibrium
(Sec.~\ref{sec:out-of-eq-optics}), the approximation related to
considering only the transitions from the core levels in optical conductivity
(Sec.~\ref{sec:core-surface-cond}), the cancellation of the equilibrium
low-frequency (static) tails at zero temperature (Sec.~\ref{sec:Cancellation-of-tails}),
the quasi-static approximation (Sec.~\ref{sec:Quasi-static-approximation}),
the damping factor $0^{+}$ (Sec.~\ref{sec:zp_study}), the transient
differential dielectric function (Sec.~\ref{sec:differential-epsilon}),
the density of states and its generalizations (Sec.~\ref{sec:Density-of-states}),
and the intra-band motion with different pump-pulse frequencies (Sec.~\ref{sec:Intra-band}).

\section{Theory\protect\label{sec:Theory}}

\subsection{Linear response in pump-probe setups\protect\label{subsec:Linear-response-in}}

In a pump-probe experimental setup, the sample is pumped using an
intense electromagnetic pulse (e.g., an IR pulse) and probed by another
pulse (e.g., an ultrashort XUV pulse) of much lower intensity (usually
just above the signal-to-noise ratio threshold). Accordingly, the
response of the system to the pump pulse is usually highly non-linear
and very complex, and requires a full-fledged quantum-mechanical theoretical
description even when the electromagnetic field is described just
classically through its time-dependent vector and scalar potentials.
On the other hand, the response of the pumped system to the probe
pulse can be safely studied up to the first order in the probe pulse,
that is within the linear response theory. This can be exploited to
greatly simplify theoretical description of pump-probe setups. In
the following, we derive and describe a comprehensive theoretical
formulation to compute the linear response of a pumped system to a
probe pulse.

The Hamiltonian that describes a system in a pump-probe setup, $\hat{H}\left(t\right)$,
can be written as 
\begin{equation}
\hat{H}\left(t\right)=\hat{H}^{\prime}\left(t\right)+\hat{H}^{\prime\prime}\left(t\right),
\end{equation}
where $\hat{H}^{\prime}\left(t\right)$ is the Hamiltonian describing
the pumped system and $\hat{H}^{\prime\prime}\left(t\right)$ is the
Hamiltonian describing the interaction between the system and the
probe pulse (the probe Hamiltonian), with respect to which we will
compute the linear response. $\hat{H}^{\prime}\left(t\right)$ is
given by the sum of the Hamiltonian describing the system at equilibrium,
$\hat{H}_{0}$, and the Hamiltonian describing the coupling between
the system and the pump pulse, $\hat{H}_{\mathrm{pu}}\left(t\right)$
(the pump Hamiltonian),
\begin{equation}
\hat{H}^{\prime}\left(t\right)=\hat{H}_{0}+\hat{H}_{\mathrm{pu}}\left(t\right).
\end{equation}
We assume that, at an initial time $t_{\mathrm{ini}}$ (e.g., $t_{\mathrm{ini}}\rightarrow-\infty$),
both pump and probe Hamiltonians vanish and the system is at equilibrium.

Given an observable that describes a type of response of the system
to the pump and probe pulses (e.g., the charge, spin, or orbital current
density, the polarization density, the magnetization density, etc.)
and the corresponding operator in the Schrödinger picture, $\hat{O}\left(t\right)$,
which can generally be time dependent, its linear response to the
probe pulse has two components. The first component arises from the
evolution in time of the state of the system under the application
of the pump and probe pulses and can be fully taken into account through
the corresponding evolution in time of the density matrix of the system.
The second component is present if there is a dependence of the operator
$\hat{O}\left(t\right)$ on the probe pulse applied to the system.

To obtain the first component, we define an interaction picture where
$\hat{H}^{\prime}\left(t\right)$ acts as the \emph{non-interacting}
Hamiltonian (with respect to the probe pulse; coupling to the pump
pulse is fully taken into account) and $\hat{H}^{\prime\prime}\left(t\right)$
acts as the \emph{interaction} Hamiltonian. Any operator $\hat{Q}\left(t\right)$
in the Schrödinger picture transforms to such interaction picture
via 
\begin{equation}
\hat{Q}_{H^{\prime}}\left(t\right)=\hat{U}^{\prime}\left(t_{\mathrm{ini}},t\right)\hat{Q}\left(t\right)\hat{U}^{\prime}\left(t,t_{\mathrm{ini}}\right),\label{eq:Hp_S_trans}
\end{equation}
where $\hat{U}^{\prime}\left(t_{1},t_{2}\right)$ is the time propagator
given in Eq.~\ref{eq:U_p} of the App.~\ref{sec:DM-linear-variation}.

As it is reported in the App.~\ref{sec:DM-linear-variation}, it
is straightforward to show that the linear variation (with respect
to probe Hamiltonian) of the many-body density matrix of the system
$\hat{\Upsilon}\left(t\right)$ (both pump and probe pulses applied)
upon the application of the probe pulse is given by
\begin{align}
 & \hat{\Upsilon}\left(t\right)-\hat{\Upsilon}^{\prime}\left(t\right)=\nonumber \\
 & -\frac{\mathrm{i}}{\hbar}\int_{t_{\mathrm{ini}}}^{t}dt^{\prime}\hat{U}^{\prime}\left(t,t_{\mathrm{ini}}\right)\left[\hat{H}_{H^{\prime}}^{\prime\prime}\left(t^{\prime}\right),\hat{\Upsilon}_{0}\right]\hat{U}^{\prime}\left(t_{\mathrm{ini}},t\right),\label{eq:DM_LR}
\end{align}
where $\hat{\Upsilon}^{\prime}\left(t\right)=\hat{U}^{\prime}\left(t,t_{\mathrm{ini}}\right)\hat{\Upsilon}_{0}\hat{U}^{\prime}\left(t_{\mathrm{ini}},t\right)$
is the many-body density matrix of the system with no probe pulse
applied and $\hat{\Upsilon}_{0}=\hat{\Upsilon}\left(t_{\mathrm{ini}}\right)$
is the equilibrium many-body density matrix of the system (nor pump
neither probe pulse applied).

The first component of the linear response, $\delta_{1}\left\langle \hat{O}\left(t\right)\right\rangle _{t}$,
comes from the variation of the density matrix and can be written
as
\begin{equation}
\delta_{1}\left\langle \hat{O}\left(t\right)\right\rangle _{t}=\mathrm{Tr}\left(\hat{O}\left(t\right)\left[\hat{\Upsilon}\left(t\right)-\hat{\Upsilon}^{\prime}\left(t\right)\right]\right),
\end{equation}
and, up to the linear order in the probe Hamiltonian, is given by

\begin{equation}
\delta_{1}\left\langle \hat{O}\left(t\right)\right\rangle _{t}=-\frac{\mathrm{i}}{\hbar}\int_{t_{\mathrm{ini}}}^{t}dt^{\prime}\left\langle \left[\hat{O}_{H^{\prime}}\left(t\right),\hat{H}_{H^{\prime}}^{\prime\prime}\left(t^{\prime}\right)\right]\right\rangle _{t_{\mathrm{ini}}},\label{eq:kubo_1st}
\end{equation}
where we used Eq.~\ref{eq:DM_LR} and $\left\langle ...\right\rangle _{t_{\mathrm{ini}}}=\mathrm{Tr}\left(...\hat{\Upsilon}_{0}\right)$
is the expectation value with respect to equilibrium many-body density
matrix of the system. Eq.~\ref{eq:kubo_1st} is formally very similar
to the Kubo formula at equilibrium.

The second component of the linear response comes from the variation
of the operator $\hat{O}\left(t\right)$ induced by the application
of the probe pulse, and is given by

\begin{equation}
\delta_{2}\left\langle \hat{O}\left(t\right)\right\rangle _{t}=\left\langle \delta\hat{O}_{H^{\prime}}\left(t\right)\right\rangle _{t_{\mathrm{ini}}},\label{eq:kubo_2nd}
\end{equation}
where $\delta\hat{O}_{H^{\prime}}\left(t\right)$ is the variation
of the operator $\hat{O}_{H^{\prime}}\left(t\right)$ up to the linear
order in the probe pulse. The total linear response to the probe pulse
is then the sum of the two components:
\begin{equation}
\delta\left\langle \hat{O}\left(t\right)\right\rangle _{t}=\delta_{1}\left\langle \hat{O}\left(t\right)\right\rangle _{t}+\delta_{2}\left\langle \hat{O}\left(t\right)\right\rangle _{t}.\label{eq:linvar_1p2}
\end{equation}

\subsection{Pumped lattice systems with quadratic Hamiltonians\protect\label{subsec:Pumped-lattice-systems}}

In this section, we consider a lattice system with an effective quadratic
(in annihilation and creation electronic operators) Hamiltonian $\hat{H}^{\prime}\left(t\right)$,

\begin{equation}
\hat{H}^{\prime}\left(t\right)=\sum_{\boldsymbol{k},n_{1},n_{2}}\hat{c}_{\boldsymbol{k},n_{1}}^{\dagger}h_{\boldsymbol{k},n_{1},n_{2}}^{\prime}\left(t\right)\hat{c}_{\boldsymbol{k},n_{2}},\label{eq:H-prime}
\end{equation}
where $\boldsymbol{k}$ is the crystal momentum that is summed over
the first Brillouin zone (FBZ), $n_{m}$ stands for all other relevant
quantum numbers (e.g., band/orbital, spin, etc.), $\hat{c}_{\boldsymbol{k},n_{m}}$
($\hat{c}_{\boldsymbol{k},n_{m}}^{\dagger}$) is the annihilation
(creation) operator of an electron in the state with quantum numbers
$\left(\boldsymbol{k},n_{m}\right)$, and $h_{\boldsymbol{k},n_{1},n_{2}}^{\prime}\left(t\right)$
is the matrix element, between the two states $\left(\boldsymbol{k},n_{1}\right)$
and $\left(\boldsymbol{k},n_{2}\right)$, of the first-quantization
(single-particle) Hamiltonian describing the sample and its coupling
to the pump pulse.

We already clarified in the previous section, given the relative low
intensity of the probe pulse inherent to its \emph{probing} nature,
that one can use the linear response theory to compute the response
of the pumped system to the probe pulse. This allows us to safely
retain only the lowest order terms in the probe pulse in the probe
Hamiltonian, $\hat{H}^{\prime\prime}\left(t\right)$. According to
this, $\hat{H}^{\prime\prime}\left(t\right)$ has the following general
form (see Eq.~\ref{eq:H-A-DG} and Refs.~\citep{schuler2021gauge,eskandari2023dynamical})
\begin{multline}
\hat{H}^{\prime\prime}\left(t\right)=\sum_{\boldsymbol{k},n_{1},n_{2}}\hat{c}_{\boldsymbol{k},n_{1}}^{\dagger}\Bigl[\frac{e}{\hbar}\boldsymbol{\mathsf{V}}_{\boldsymbol{k},n_{1},n_{2}}\left(t\right)\cdot\boldsymbol{A}_{\mathrm{pr}}\left(t\right)\\
+e\boldsymbol{\mathsf{D}}_{\boldsymbol{k},n_{1},n_{2}}\left(t\right)\cdot\boldsymbol{E}_{\mathrm{pr}}\left(t\right)\Bigr]\hat{c}_{\boldsymbol{k},n_{2}},\label{eq:H_second}
\end{multline}
where $\cdot$ is the inner product of vectors, $\boldsymbol{A}_{\mathrm{pr}}\left(t\right)$
and $\boldsymbol{E}_{\mathrm{pr}}\left(t\right)=-\partial_{t}\boldsymbol{A}_{\mathrm{pr}}\left(t\right)$
are the probe-pulse vector potential and electric field, respectively,
which we assume to be linearly polarized, $-e$ is the charge of electron,
and the time- and momentum-dependent vectorial matrix elements $\boldsymbol{\mathsf{V}}_{\boldsymbol{k},n_{1},n_{2}}\left(t\right)$
and $\boldsymbol{\mathsf{D}}_{\boldsymbol{k},n_{1},n_{2}}\left(t\right)$
determine, in general, how the vector potential and electric field
couple to the sample, respectively. Later, we will give their explicit
expressions in the dipole gauge (see Eqs.~\ref{eq:Dnnk} and \ref{eq:Vk}).
We have assumed that the wavelength of the probe pulse is not shorter
than the XUV one, which is hundreds of times larger than any lattice
constant in \emph{real} materials, so that one can consider the electromagnetic
field practically as uniform. The cases with much shorter wavelengths
are beyond the scope of this work.

Then, for a generic observable $\hat{O}\left(t\right)=\sum_{\boldsymbol{k},n_{1},n_{2}}\hat{c}_{\boldsymbol{k},n_{1}}^{\dagger}O_{\boldsymbol{k},n_{1},n_{2}}\left(t\right)\hat{c}_{\boldsymbol{k},n_{2}}$,
the first component of the linear response to the probe pulse is obtained
from Eq.~\ref{eq:kubo_1st} as
\begin{widetext}
\begin{multline}
\delta_{1}\left\langle \hat{O}\left(t\right)\right\rangle _{t}=-\frac{\mathrm{i}}{\hbar}\sum_{n_{1},n_{2},n_{3},n_{4}}\sum_{\boldsymbol{k}}O_{\boldsymbol{k},n_{1},n_{2}}\left(t\right)\int_{t_{\mathrm{ini}}}^{t}dt^{\prime}\left(\frac{e}{\hbar}\boldsymbol{\mathsf{V}}_{\boldsymbol{k},n_{3},n_{4}}\left(t^{\prime}\right)\cdot\boldsymbol{A}_{\mathrm{pr}}\left(t^{\prime}\right)+e\boldsymbol{\mathsf{D}}_{\boldsymbol{k},n_{3},n_{4}}\left(t^{\prime}\right)\cdot\boldsymbol{E}_{\mathrm{pr}}\left(t^{\prime}\right)\right)\times\\
\times\left\langle \left[\hat{c}_{H^{\prime},\boldsymbol{k},n_{1}}^{\dagger}\left(t\right)\hat{c}_{H^{\prime},\boldsymbol{k},n_{2}}\left(t\right),\hat{c}_{H^{\prime},\boldsymbol{k},n_{3}}^{\dagger}\left(t^{\prime}\right)\hat{c}_{H^{\prime},\boldsymbol{k},n_{4}}\left(t^{\prime}\right)\right]\right\rangle _{t_{\mathrm{ini}}}.
\end{multline}
\end{widetext}

As a mathematical simplification, without any loss of generality,
we consider the electric field of the probe pulse to be a Dirac delta
function applied at the time $t_{\mathrm{pr}}$, the probe time, 
\begin{equation}
\boldsymbol{E}_{\mathrm{pr}}\left(t\right)=\boldsymbol{E}_{\mathrm{pr}}\delta\left(t-t_{\mathrm{pr}}\right),\label{eq:Epr}
\end{equation}
and, consequently, we have
\begin{equation}
\boldsymbol{A}_{\mathrm{pr}}\left(t\right)=-\boldsymbol{E}_{\mathrm{pr}}\theta\left(t-t_{\mathrm{pr}}\right).\label{eq:Apr}
\end{equation}
A Dirac-delta shape for the probe electric field (as in Green\textquoteright s
function theory for differential equations) does not imply any restriction
in the shape of the actual probe pulse, as any probe pulse can be
expanded in terms of Dirac-delta functions, and the final result (the
expression of the optical conductivity, Eqs.~\ref{eq:sigma-1-P-t}
and \ref{eq:sigma-2-P-t}) does not depend on this choice. A detailed
derivation is given in the App.~\ref{sec:out-of-eq-optics}. Hereafter,
we assume that $t_{\mathrm{pr}}\geq t_{\mathrm{ini}}$.

The first component of the linear response to such a probe reads as
\begin{widetext}
\begin{multline}
\delta_{1}\left\langle \hat{O}\left(t\right)\right\rangle _{t}=-\frac{\mathrm{i}}{\hbar}\sum_{n_{1},n_{2},n_{3},n_{4}}\sum_{\boldsymbol{k}}O_{\boldsymbol{k},n_{1},n_{2}}\left(t\right)\int_{t_{\mathrm{ini}}}^{t}dt^{\prime}\left(-\frac{e}{\hbar}\boldsymbol{\mathsf{V}}_{\boldsymbol{k},n_{3},n_{4}}\left(t^{\prime}\right)\theta\left(t^{\prime}-t_{\mathrm{pr}}\right)+e\boldsymbol{\mathsf{D}}_{\boldsymbol{k},n_{3},n_{4}}\left(t_{\mathrm{pr}}\right)\delta\left(t^{\prime}-t_{\mathrm{pr}}\right)\right)\cdot\boldsymbol{E}_{\mathrm{pr}}\times\\
\times\left\langle \left[\hat{c}_{H^{\prime},\boldsymbol{k},n_{1}}^{\dagger}\left(t\right)\hat{c}_{H^{\prime},\boldsymbol{k},n_{2}}\left(t\right),\hat{c}_{H^{\prime},\boldsymbol{k},n_{3}}^{\dagger}\left(t^{\prime}\right)\hat{c}_{H^{\prime},\boldsymbol{k},n_{4}}\left(t^{\prime}\right)\right]\right\rangle _{t_{\mathrm{ini}}}.\label{eq:DO1-Epr}
\end{multline}
\end{widetext}

The retarded response function of the system, $\boldsymbol{\chi}\left(t,t^{\prime}\right)$,
is defined through

\begin{equation}
\delta\left\langle \hat{O}\left(t\right)\right\rangle _{t}=\int_{t_{\mathrm{ini}}}^{\infty}dt^{\prime}\boldsymbol{\chi}\left(t,t^{\prime}\right)\cdot\boldsymbol{E}_{\mathrm{pr}}\left(t^{\prime}\right).\label{eq:resp_gen_form}
\end{equation}
Obviously, if $\hat{O}$ is a vector (scalar), $\boldsymbol{\chi}$
is a rank 2 (rank 1) tensor. Since the linear response has two components
(see Eq.~\ref{eq:linvar_1p2}), the response function too can be
written as the sum of two components, $\boldsymbol{\chi}\left(t,t^{\prime}\right)=\boldsymbol{\chi}_{1}\left(t,t^{\prime}\right)+\boldsymbol{\chi}_{2}\left(t,t^{\prime}\right)$,
where $\boldsymbol{\chi}_{1\left(2\right)}\left(t,t^{\prime}\right)$
corresponds to the first (second) component of the linear response
to the probe pulse. Given the expression chosen for the probe pulse
(Eq.~\ref{eq:Epr}), Eq.~\ref{eq:resp_gen_form} simplifies to
\begin{equation}
\delta_{1,2}\left\langle \hat{O}\left(t\right)\right\rangle _{t}=\boldsymbol{\chi}_{1,2}\left(t,t_{\mathrm{pr}}\right)\cdot\boldsymbol{E}_{\mathrm{pr}}.\label{eq:O.X.Epr}
\end{equation}
Comparing Eq.~\ref{eq:O.X.Epr} with Eq.~\ref{eq:DO1-Epr} and considering
the arbitrariness of $\boldsymbol{E}_{\mathrm{pr}}$, we get the following
expression for the first component of the response function
\begin{widetext}
\begin{multline}
\boldsymbol{\chi}_{1}\left(t,t_{\mathrm{pr}}\right)=-\frac{\mathrm{i}}{\hbar}\theta\left(t-t_{\mathrm{pr}}\right)\\
\sum_{n_{1},n_{2},n_{3},n_{4}}\sum_{\boldsymbol{k}}O_{\boldsymbol{k},n_{1},n_{2}}\left(t\right)\int_{t_{\mathrm{ini}}}^{t}dt^{\prime}\left(-\frac{e}{\hbar}\boldsymbol{\mathsf{V}}_{\boldsymbol{k},n_{3},n_{4}}\left(t^{\prime}\right)\theta\left(t^{\prime}-t_{\mathrm{pr}}\right)+e\boldsymbol{\mathsf{D}}_{\boldsymbol{k},n_{3},n_{4}}\left(t_{\mathrm{pr}}\right)\delta\left(t^{\prime}-t_{\mathrm{pr}}\right)\right)\times\\
\times\left\langle \left[\hat{c}_{H^{\prime},\boldsymbol{k},n_{1}}^{\dagger}\left(t\right)\hat{c}_{H^{\prime},\boldsymbol{k},n_{2}}\left(t\right),\hat{c}_{H^{\prime},\boldsymbol{k},n_{3}}^{\dagger}\left(t^{\prime}\right)\hat{c}_{H^{\prime},\boldsymbol{k},n_{4}}\left(t^{\prime}\right)\right]\right\rangle _{t_{\mathrm{ini}}}.\label{eq:X1-cH}
\end{multline}
\end{widetext}

For the second component of the response function, we need the variation
of the operator $\hat{O}\left(t\right)$ induced by the application
of the probe pulse up to the linear order. We assume that $\hat{O}\left(t\right)$
depends on the applied pulses only through their vector potentials,
as it is the case for the current operator. Then, considering the
expression chosen for the probe pulse, Eq.~\ref{eq:Epr}, we can
write
\begin{multline}
\delta_{2}\left\langle \hat{O}\left(t\right)\right\rangle _{t}=\sum_{n_{1},n_{2}}\sum_{\boldsymbol{k}}\frac{\delta O_{\boldsymbol{k},n_{1},n_{2}}\left(t\right)}{\delta\boldsymbol{A}\left(t\right)}\cdot\boldsymbol{A}_{\mathrm{pr}}\left(t\right)\\
\left\langle \hat{c}_{H^{\prime},\boldsymbol{k},n_{1}}^{\dagger}\left(t\right)\hat{c}_{H^{\prime},\boldsymbol{k},n_{2}}\left(t\right)\right\rangle _{t_{\mathrm{ini}}}\\
=-\theta\left(t-t_{\mathrm{pr}}\right)\sum_{n_{1},n_{2}}\sum_{\boldsymbol{k}}\frac{\delta O_{\boldsymbol{k},n_{1},n_{2}}\left(t\right)}{\delta\boldsymbol{A}\left(t\right)}\cdot\boldsymbol{E}_{\mathrm{pr}}\\
\left\langle \hat{c}_{H^{\prime},\boldsymbol{k},n_{1}}^{\dagger}\left(t\right)\hat{c}_{H^{\prime},\boldsymbol{k},n_{2}}\left(t\right)\right\rangle _{t_{\mathrm{ini}}}.\label{eq:DO2-Epr}
\end{multline}
Comparing Eq.~\ref{eq:DO2-Epr} with Eq.~\ref{eq:O.X.Epr}, we get
\begin{multline}
\boldsymbol{\chi}_{2}\left(t,t_{\mathrm{pr}}\right)=-\theta\left(t-t_{\mathrm{pr}}\right)\sum_{n_{1},n_{2}}\sum_{\boldsymbol{k}}\frac{\delta O_{\boldsymbol{k},n_{1},n_{2}}\left(t\right)}{\delta\boldsymbol{A}\left(t\right)}\\
\left\langle \hat{c}_{H^{\prime},\boldsymbol{k},n_{1}}^{\dagger}\left(t\right)\hat{c}_{H^{\prime},\boldsymbol{k},n_{2}}\left(t\right)\right\rangle _{t_{\mathrm{ini}}}.\label{eq:X2-cH}
\end{multline}

Then, to evaluate $\boldsymbol{\chi}\left(t,t_{\mathrm{pr}}\right)$,
one needs to compute the time evolution of the creation and annihilation
operators, $\hat{c}_{H^{\prime},\boldsymbol{k},n_{m}}^{\dagger}\left(t\right)$
and $\hat{c}_{H^{\prime},\boldsymbol{k},n_{m}}\left(t\right)$, appearing
in Eqs.~\ref{eq:X1-cH} and \ref{eq:X2-cH}. In a Heisenberg-like
fashion, the dynamics of these operators encodes all of the effects
of applying the pump pulse to the system. According to what we have
recently shown in Refs.~\citep{inzani2022field,eskandari2023dynamical},
using the Dynamical Projective Operatorial Approach (DPOA), such operators
can be projected on their equilibrium counterparts, 
\begin{equation}
\hat{c}_{H^{\prime},\boldsymbol{k},n_{1}}\left(t\right)=\sum_{n_{2}}P_{\boldsymbol{k},n_{1},n_{2}}\left(t\right)\hat{c}_{\boldsymbol{k},n_{2}},
\end{equation}
and the projection coefficients, $P_{\boldsymbol{k},n_{1},n_{2}}\left(t\right)$,
are obtained from the following equation of motion
\begin{equation}
\mathrm{i}\hbar\partial_{t}P_{\boldsymbol{k},n_{1},n_{2}}\left(t\right)=\sum_{n_{3}}h_{\boldsymbol{k},n_{1},n_{3}}^{\prime}\left(t\right)P_{\boldsymbol{k},n_{3},n_{2}}\left(t\right),\label{eq:P-dyn-gen}
\end{equation}
with the initial condition $P_{\boldsymbol{k},n_{1},n_{2}}\left(t_{\mathrm{ini}}\right)=\delta_{n_{1},n_{2}}$.
Using such projection coefficients and assuming that the equilibrium
Hamiltonian, $\hat{H}_{0}$, is diagonal in the basis we have chosen,
i.e., $\hat{H}_{0}=\sum_{\boldsymbol{k},n}\varepsilon_{\boldsymbol{k},n}\hat{c}_{\boldsymbol{k},n}^{\dagger}\hat{c}_{\boldsymbol{k},n},$
one can compute the two components of the response function as
\begin{widetext}
\begin{multline}
\boldsymbol{\chi}_{1}\left(t,t_{\mathrm{pr}}\right)=-\frac{\mathrm{i}}{\hbar}\theta\left(t-t_{\mathrm{pr}}\right)\\
\sum_{n_{1},n_{2},n_{3},n_{4}}\sum_{n_{1}^{\prime},n_{2}^{\prime}}\sum_{\boldsymbol{k}}\hat{O}_{\boldsymbol{k},n_{1},n_{2}}\left(t\right)\int_{t_{\mathrm{ini}}}^{t}dt^{\prime}\left(-\frac{e}{\hbar}\boldsymbol{\mathsf{V}}_{\boldsymbol{k},n_{3},n_{4}}\left(t^{\prime}\right)\theta\left(t^{\prime}-t_{\mathrm{pr}}\right)+e\boldsymbol{\mathsf{D}}_{\boldsymbol{k},n_{3},n_{4}}\left(t_{\mathrm{pr}}\right)\delta\left(t^{\prime}-t_{\mathrm{pr}}\right)\right)\times\\
\times P_{\boldsymbol{k},n_{1},n_{1}^{\prime}}^{\star}\left(t\right)P_{\boldsymbol{k},n_{2},n_{2}^{\prime}}\left(t\right)P_{\boldsymbol{k},n_{3},n_{2}^{\prime}}^{\star}\left(t^{\prime}\right)P_{\boldsymbol{k},n_{4},n_{1}^{\prime}}\left(t^{\prime}\right)\left(f_{\boldsymbol{k},n_{1}^{\prime}}-f_{\boldsymbol{k},n_{2}^{\prime}}\right),\label{eq:khi_1}
\end{multline}
\end{widetext}

and
\begin{multline}
\boldsymbol{\chi}_{2}\left(t,t_{\mathrm{pr}}\right)=-\theta\left(t-t_{\mathrm{pr}}\right)\\
\sum_{n_{1},n_{2},n_{3}}\sum_{\boldsymbol{k}}\frac{\delta O_{\boldsymbol{k},n_{1},n_{2}}\left(t\right)}{\delta\boldsymbol{A}\left(t\right)}P_{\boldsymbol{k},n_{1},n_{3}}^{\star}\left(t\right)P_{\boldsymbol{k},n_{2},n_{3}}\left(t\right)f_{\boldsymbol{k},n_{3}},\label{eq:eq:khi_2}
\end{multline}
where we have used $\left\langle \hat{c}_{\boldsymbol{k},n_{1}}^{\dagger}\hat{c}_{\boldsymbol{k},n_{2}}\right\rangle _{t_{\mathrm{ini}}}=\delta_{n_{1},n_{2}}f_{\boldsymbol{k},n_{1}}$
and $\left\langle \left[c_{\boldsymbol{k},n_{1}}^{\dagger}c_{\boldsymbol{k},n_{2}},c_{\boldsymbol{k},n_{3}}^{\dagger}c_{\boldsymbol{k},n_{4}}\right]\right\rangle _{t_{\mathrm{ini}}}=\delta_{n_{1},n_{4}}\delta_{n_{2},n_{3}}\left(f_{\boldsymbol{k},n_{1}}-f_{\boldsymbol{k},n_{2}}\right)$.
$f_{\boldsymbol{k},n}=\frac{1}{\mathrm{e}^{\left(\varepsilon_{\boldsymbol{k},n}-\mu\right)/k_{\mathrm{B}}T}+1}$
is the Fermi function that determines the equilibrium distribution
of electrons in which $k_{\mathrm{B}}$ is the Boltzmann constant,
$T$ is the temperature, and $\mu$ is the chemical potential.

\subsection{Optical conductivity\protect\label{subsec:Optical-conductivity}}

For a lattice system, in the dipole gauge, the matter-field coupling
Hamiltonian, given a generic electromagnetic pulse, reads as (see
Refs.~\citep{schuler2021gauge,eskandari2023dynamical})

\begin{multline}
\hat{H}\left\{ \boldsymbol{A}\left(t\right),\boldsymbol{E}\left(t\right)\right\} =\sum_{\nu_{1},\nu_{2},\boldsymbol{k}}\hat{c}_{\boldsymbol{k},\nu_{1}}^{\dagger}\tilde{T}_{\boldsymbol{k}+\frac{e}{\hbar}\boldsymbol{A}\left(t\right),\nu_{1},\nu_{2}}\hat{c}_{\boldsymbol{k},\nu_{2}}\\
+e\boldsymbol{E}\left(t\right)\cdot\sum_{\nu_{1}\nu_{2}\boldsymbol{k}}\hat{c}_{\boldsymbol{k}\nu_{1}}^{\dagger}\tilde{\boldsymbol{D}}_{\boldsymbol{k}+\frac{e}{\hbar}\boldsymbol{A}\left(t\right),\nu_{1}\nu_{2}}\hat{c}_{\boldsymbol{k}\nu_{2}},\label{eq:H-A-DG}
\end{multline}
where $\boldsymbol{A}\left(t\right)$ is the vector potential and
$\boldsymbol{E}\left(t\right)=-\partial_{t}\boldsymbol{A}\left(t\right)$
is the electric field. $\nu_{1}$ and $\nu_{2}$ are sets of quantum
numbers identifying orthogonal localized states (e.g., the maximally
localized Wannier states). We neglect the coupling of the spin magnetic
moments to the magnetic field of the applied pulses. $\tilde{T}_{\boldsymbol{k},\nu_{1}\nu_{2}}$
is the matrix element of the equilibrium Hamiltonian, i.e., $\hat{H}_{0}=\sum_{\nu_{1}\nu_{2}\boldsymbol{k}}\hat{c}_{\boldsymbol{k}\nu_{1}}^{\dagger}\tilde{T}_{\boldsymbol{k},\nu_{1}\nu_{2}}\hat{c}_{\boldsymbol{k}\nu_{2}}$,
and $\tilde{\boldsymbol{D}}_{\boldsymbol{k},\nu_{1}\nu_{2}}$ is the
matrix element of the local dipole in the reciprocal space. In the
absence of the probe pulse, the electric field and the vector potential
are just those of the pump pulse, i.e., $\boldsymbol{A}_{\mathrm{pu}}\left(t\right)$
and $\boldsymbol{E}_{\mathrm{pu}}\left(t\right)$, respectively. Consequently,
the Hamiltonian of the pumped system is given by $\hat{H}^{\prime}\left(t\right)=\hat{H}\left\{ \boldsymbol{A}_{\mathrm{pu}}\left(t\right),\boldsymbol{E}_{\mathrm{pu}}\left(t\right)\right\} $.

Instead of working in the localized-state basis ($\nu$ indexed),
we prefer to perform the calculations in the basis which diagonalizes
the equilibrium Hamiltonian ($n$ indexed). Using the diagonalization
matrix $\boldsymbol{\Omega}_{\boldsymbol{k}}$,

\begin{equation}
\sum_{\nu_{1}\nu_{2}}\Omega_{\boldsymbol{k}\nu_{1}n_{1}}^{*}\tilde{T}_{\boldsymbol{k},\nu_{1}\nu_{2}}\Omega_{\boldsymbol{k}\nu_{2}n_{2}}=\varepsilon_{\boldsymbol{k},n}\delta_{n_{1}n_{2}},
\end{equation}
we can rewrite $\hat{H}^{\prime}\left(t\right)$ as

\begin{align}
 & \hat{H}^{\prime}\left(t\right)=\sum_{n_{1}n_{2}\boldsymbol{k}}\hat{c}_{\boldsymbol{k}n_{1}}^{\dagger}T_{\boldsymbol{k},n_{1}n_{2}}\left(t\right)\hat{c}_{\boldsymbol{k}n_{2}}\nonumber \\
 & +e\boldsymbol{E}_{\mathrm{pu}}\left(t\right)\cdot\sum_{n_{1}n_{2}\boldsymbol{k}}\hat{c}_{\boldsymbol{k}n_{1}}^{\dagger}\boldsymbol{D}_{\boldsymbol{k}.n_{1}n_{2}}\left(t\right)\hat{c}_{\boldsymbol{k}n_{2}},\label{eq:H_prime}
\end{align}
where $\hat{c}_{\boldsymbol{k}n}^{\dagger}=\sum_{\nu}\hat{c}_{\boldsymbol{k}\nu}^{\dagger}\Omega_{\boldsymbol{k}\nu n}$.
Moreover, $T_{\boldsymbol{k},n_{1}n_{2}}\left(t\right)=T_{\boldsymbol{k},n_{1}n_{2}}\left\{ \boldsymbol{A}_{\mathrm{pu}}\left(t\right)\right\} $
in which
\begin{equation}
T_{\boldsymbol{k},n_{1}n_{2}}\left\{ \boldsymbol{A}\left(t\right)\right\} =\sum_{\nu_{1}\nu_{2}}\Omega_{\boldsymbol{k}\nu_{1}n_{1}}^{*}\tilde{T}_{\boldsymbol{k}+\frac{e}{\hbar}\boldsymbol{A}\left(t\right).\nu_{1}\nu_{2}}\Omega_{\boldsymbol{k}\nu_{2}n_{2}},
\end{equation}
 and $\boldsymbol{D}_{\boldsymbol{k},n_{1}n_{2}}\left(t\right)=\boldsymbol{D}_{\boldsymbol{k},n_{1}n_{2}}\left\{ \boldsymbol{A}_{\mathrm{pu}}\left(t\right)\right\} $
in which
\begin{equation}
\boldsymbol{D}_{\boldsymbol{k},n_{1}n_{2}}\left\{ \boldsymbol{A}\left(t\right)\right\} =\sum_{\nu_{1}\nu_{2}}\Omega_{\boldsymbol{k}\nu_{1}n_{1}}^{*}\tilde{\boldsymbol{D}}_{\boldsymbol{k}+\frac{e}{\hbar}\boldsymbol{A}\left(t\right),\nu_{1}\nu_{2}}\Omega_{\boldsymbol{k}\nu_{2}n_{2}}.\label{eq:Dnnk}
\end{equation}

In the presence of both pump and probe pulses, the actual applied
fields are given by $\boldsymbol{A}_{\mathrm{pu}}\left(t\right)+\boldsymbol{A}_{\mathrm{pr}}\left(t\right)$
and $\boldsymbol{E}_{\mathrm{pu}}\left(t\right)+\boldsymbol{E}_{\mathrm{pr}}\left(t\right)$.
As discussed above, the probe pulse intensity is relatively weak and
we are considering it within the linear response theory. Therefore,
in order to obtain the coupling Hamiltonian between the system and
the probe pulse, $\hat{H}^{\prime\prime}\left(t\right)$, we expand
the full Hamiltonian, $\hat{H}\left\{ \boldsymbol{A}_{\mathrm{pu}}\left(t\right)+\boldsymbol{A}_{\mathrm{pr}}\left(t\right),\boldsymbol{E}_{\mathrm{pu}}\left(t\right)+\boldsymbol{E}_{\mathrm{pr}}\left(t\right)\right\} $,
up to the linear order in the probe-pulse fields and subtract $\hat{H}^{\prime}\left(t\right)$.
It is worth noticing that we do not need to expand in terms of $\boldsymbol{E}_{\mathrm{pr}}\left(t\right)$
as it only appears already at the linear order. Following this procedure,
it's straightforward to show that $\hat{H}^{\prime\prime}\left(t\right)$
is of the form given in Eq.~\ref{eq:H_second}, with $\boldsymbol{\mathsf{D}}_{\boldsymbol{k},n_{1}n_{2}}\left(t\right)=\boldsymbol{D}_{\boldsymbol{k},n_{1}n_{2}}\left\{ \boldsymbol{A}_{\mathrm{pu}}\left(t\right)\right\} $
(see Eq.~\ref{eq:Dnnk}), and $\boldsymbol{\mathsf{V}}_{\boldsymbol{k},n_{1}n_{2}}\left(t\right)=\boldsymbol{\mathsf{V}}_{\boldsymbol{k},n_{1}n_{2}}\left\{ \boldsymbol{A}_{\mathrm{pu}}\left(t\right),\boldsymbol{E}_{\mathrm{pu}}\left(t\right)\right\} $,
where
\begin{multline}
\boldsymbol{\mathsf{V}}_{\boldsymbol{k},n_{1}n_{2}}\left\{ \boldsymbol{A}\left(t\right),\boldsymbol{E}\left(t\right)\right\} =\\
\boldsymbol{\eta}_{\boldsymbol{k},n_{1}n_{2}}\left\{ \boldsymbol{A}\left(t\right)\right\} +e\boldsymbol{\Lambda}_{\boldsymbol{k},n_{1}n_{2}}\left\{ \boldsymbol{A}\left(t\right)\right\} \cdot\boldsymbol{E}\left(t\right),\label{eq:Vk}
\end{multline}
in which
\begin{equation}
\boldsymbol{\eta}_{\boldsymbol{k},n_{1}n_{2}}\left\{ \boldsymbol{A}\left(t\right)\right\} =\sum_{\nu_{1}\nu_{2}}\Omega_{\boldsymbol{k}\nu_{1}n_{1}}^{*}\left[\boldsymbol{\nabla}_{\boldsymbol{k}}\tilde{T}_{\boldsymbol{k}+\frac{e}{\hbar}\boldsymbol{A}\left(t\right),\nu_{1}\nu_{2}}\right]\Omega_{\boldsymbol{k}\nu_{2}n_{2}},
\end{equation}
is proportional to the velocity at the crystal momentum $\boldsymbol{k}+\frac{e}{\hbar}\boldsymbol{A}\left(t\right)$
transformed to the equilibrium diagonal basis at the crystal momentum
$\boldsymbol{k}$. Moreover,
\begin{equation}
\boldsymbol{\Lambda}_{\boldsymbol{k},n_{1}n_{2}}\left\{ \boldsymbol{A}\left(t\right)\right\} =\sum_{\nu_{1}\nu_{2}}\Omega_{\boldsymbol{k}\nu_{1}n_{1}}^{*}\left[\boldsymbol{\nabla}_{\boldsymbol{k}}\tilde{\boldsymbol{D}}_{\boldsymbol{k}+\frac{e}{\hbar}\boldsymbol{A}\left(t\right),\nu_{1}\nu_{2}}\right]\Omega_{\boldsymbol{k}\nu_{2}n_{2}},\label{eq:VD}
\end{equation}
is a rank 2 tensor with the first index in momentum space coordinates
(coming from $\boldsymbol{\nabla}_{\boldsymbol{k}}$) and the second
index in direct space ones (coming from $\tilde{\boldsymbol{D}}_{\boldsymbol{k},\nu_{1}\nu_{2}}$).
This latter is the one saturated by the scalar product with the electric
field.

The optical conductivity is the response function describing how an
electric current is induced by the probe pulse. The zero-momentum
particle current (i.e., the particle current integrated over the pulse-interacting
portion of the volume of the sample), in the presence of the pump
field and in the dipole gauge is given by (see Refs.~\citep{schuler2021gauge,eskandari2023dynamical})
$\hat{\boldsymbol{J}}\left(t\right)=\hat{\boldsymbol{J}}\left\{ \boldsymbol{A}_{\mathrm{pu}}\left(t\right)\right\} $
where
\begin{equation}
\hat{\boldsymbol{J}}\left\{ \boldsymbol{A}\left(t\right)\right\} =\sum_{n_{1}n_{2}\boldsymbol{k}}\hat{c}_{\boldsymbol{k}n_{1}}^{\dagger}\boldsymbol{J}_{\boldsymbol{k},n_{1}n_{2}}\left\{ \boldsymbol{A}\left(t\right)\right\} \hat{c}_{\boldsymbol{k}n_{2}},
\end{equation}
in which

\begin{multline}
\boldsymbol{J}_{\boldsymbol{k},n_{1}n_{2}}\left\{ \boldsymbol{A}\left(t\right)\right\} =\frac{1}{\hbar}\boldsymbol{\eta}_{\boldsymbol{k},n_{1}n_{2}}\left\{ \boldsymbol{A}\left(t\right)\right\} \\
-\frac{\mathrm{i}}{\hbar}\left[\boldsymbol{D}_{\boldsymbol{k}}\left\{ \boldsymbol{A}\left(t\right)\right\} ,\boldsymbol{T}_{\boldsymbol{k}}\left\{ \boldsymbol{A}\left(t\right)\right\} \right]_{n_{1}n_{2}},
\end{multline}
where $\left[\varPhi,\varPsi\right]_{n_{1}n_{2}}=\sum_{n^{\prime}}\left(\varPhi_{n_{1}n^{\prime}}\varPsi_{n^{\prime}n_{2}}-\varPsi_{n_{1}n^{\prime}}\varPhi_{n^{\prime}n_{2}}\right)$.

The macroscopic particle current, which is the averaged microscopic
particle current over a macroscopic number of unit cells, is just
the zero-momentum current divided by the pulse-interacting portion
of the volume of the sample, $\mathcal{V}$. Therefore, the macroscopic
electric current, or simply the electric current, is $-e\hat{\boldsymbol{J}}\left(t\right)/\mathcal{V}$.
Using Eq.~\ref{eq:khi_1}, with $\hat{O}\left(t\right)=-e\hat{\boldsymbol{J}}\left(t\right)/\mathcal{V}$,
the first component of optical conductivity is obtained as
\begin{widetext}
\begin{multline}
\boldsymbol{\sigma}_{1}\left(t,t_{\mathrm{pr}}\right)=\frac{\mathrm{i}e}{\hbar\mathcal{V}}\theta\left(t-t_{\mathrm{pr}}\right)\\
\sum_{\boldsymbol{k}}\sum_{n_{1}n_{2}n_{3}n_{4}}\sum_{n_{1}^{\prime}n_{2}^{\prime}}\boldsymbol{J}_{\boldsymbol{k},n_{1}n_{2}}\left(t\right)\int_{t_{\mathrm{ini}}}^{t}dt^{\prime}\left(-\frac{e}{\hbar}\boldsymbol{\mathsf{V}}_{\boldsymbol{k},n_{3}n_{4}}\left(t^{\prime}\right)\theta\left(t^{\prime}-t_{\mathrm{pr}}\right)+e\boldsymbol{\mathsf{D}}_{\boldsymbol{k},n_{3}n_{4}}\left(t^{\prime}\right)\delta\left(t^{\prime}-t_{\mathrm{pr}}\right)\right)\times\\
\times P_{\boldsymbol{k}n_{1}n_{1}^{\prime}}^{\star}\left(t\right)P_{\boldsymbol{k}n_{2}n_{2}^{\prime}}\left(t\right)P_{\boldsymbol{k}n_{3}n_{2}^{\prime}}^{\star}\left(t^{\prime}\right)P_{\boldsymbol{k}n_{4}n_{1}^{\prime}}\left(t^{\prime}\right)\left(f_{\boldsymbol{k}n_{1}^{\prime}}-f_{\boldsymbol{k}n_{2}^{\prime}}\right).\label{eq:sigma-1-P-t}
\end{multline}
\end{widetext}

For the second component of the optical conductivity, we have

\begin{align}
 & \boldsymbol{\sigma}_{2}\left(t,t_{\mathrm{pr}}\right)=\frac{e}{\mathcal{V}}\theta\left(t-t_{\mathrm{pr}}\right)\sum_{n_{1}n_{2}\boldsymbol{k}}\frac{\delta\boldsymbol{J}_{\boldsymbol{k},n_{1}n_{2}}\left(t\right)}{\delta\boldsymbol{A}\left(t\right)}N_{\boldsymbol{k}n_{1}n_{2}}\left(t\right),\label{eq:sigma-2-P-t}
\end{align}
where
\begin{equation}
N_{\boldsymbol{k}n_{1}n_{2}}\left(t\right)=\sum_{n^{\prime}}P_{\boldsymbol{k}n_{1}n^{\prime}}^{\star}\left(t\right)P_{\boldsymbol{k}n_{2}n^{\prime}}\left(t\right)f_{\boldsymbol{k}n^{\prime}},
\end{equation}
and
\begin{multline}
\frac{\delta\boldsymbol{J}_{\boldsymbol{k},n_{1}n_{2}}\left(t\right)}{\delta\boldsymbol{A}\left(t\right)}=\frac{e}{\hbar^{2}}\boldsymbol{\xi}_{\boldsymbol{k},n_{1}n_{2}}\left(t\right)\\
-\frac{\mathrm{i}e}{\hbar^{2}}\left[\boldsymbol{\Lambda}_{\boldsymbol{k}}\left(t\right),T_{\boldsymbol{k}}\left(t\right)\right]_{n_{1}n_{2}}-\frac{\mathrm{i}e}{\hbar^{2}}\left[\boldsymbol{D}_{\boldsymbol{k}}\left(t\right),\boldsymbol{\eta}_{\boldsymbol{k}}\left(t\right)\right]_{n_{1}n_{2}},
\end{multline}
in which $\boldsymbol{\Lambda}_{\boldsymbol{k}}\left(t\right)=\boldsymbol{\Lambda}_{\boldsymbol{k}}\left\{ \boldsymbol{A}_{\mathrm{pu}}\left(t\right)\right\} $,
$\boldsymbol{\eta}_{\boldsymbol{k},n_{1}n_{2}}\left(t\right)=\boldsymbol{\eta}_{\boldsymbol{k},n_{1}n_{2}}\left\{ \boldsymbol{A}_{\mathrm{pu}}\left(t\right)\right\} $,
and $\boldsymbol{\xi}_{\boldsymbol{k},n_{1}n_{2}}\left(t\right)=\boldsymbol{\xi}_{\boldsymbol{k},n_{1}n_{2}}\left\{ \boldsymbol{A}_{\mathrm{pu}}\left(t\right)\right\} $
where
\begin{equation}
\boldsymbol{\xi}_{\boldsymbol{k},n_{1}n_{2}}\left\{ \boldsymbol{A}\left(t\right)\right\} =\sum_{\nu_{1}\nu_{2}}\Omega_{\boldsymbol{k}\nu_{1}n_{1}}^{*}\left[\nabla_{\boldsymbol{k}}^{2}\tilde{T}_{\boldsymbol{k}+\frac{e}{\hbar}\boldsymbol{A}\left(t\right),\nu_{1}\nu_{2}}\right]\Omega_{\boldsymbol{k}\nu_{2}n_{2}},
\end{equation}
is proportional to the inverse-mass tensor at the crystal momentum
$\boldsymbol{k}+\frac{e}{\hbar}\boldsymbol{A}\left(t\right)$ transformed
to the equilibrium diagonal basis at the crystal momentum $\boldsymbol{k}$.

The total optical conductivity is thus obtained by adding up its two
components as,

\begin{equation}
\boldsymbol{\sigma}\left(t,t_{\mathrm{pr}}\right)=\boldsymbol{\sigma}_{1}\left(t,t_{\mathrm{pr}}\right)+\boldsymbol{\sigma}_{2}\left(t,t_{\mathrm{pr}}\right).\label{eq:sigma-P-t}
\end{equation}

We choose the center of the pump pulse envelope as the origin of the
time axis. Accordingly, $t_{\mathrm{pr}}$ is just the time delay
between pump and probe pulses. One is usually interested in the optical
conductivity in terms of frequency and time delay, which is obtained
by performing a Fourier transformation with respect to $t-t_{\mathrm{pr}}$,
i.e., 
\begin{equation}
\boldsymbol{\sigma}\left(\omega,t_{\mathrm{pr}}\right)=\int_{-\infty}^{+\infty}\mathrm{e}^{\mathrm{i}\left(\omega+\mathrm{i}0^{+}\right)\left(t-t_{\mathrm{pr}}\right)}\boldsymbol{\sigma}\left(t,t_{\mathrm{pr}}\right)dt,\label{eq:FT-sigma}
\end{equation}
where $0^{+}$ is a small damping factor.

\subsection{Reflectivity and Absorption\protect\label{subsec:Reflectivity-and-Absorption}}

In this section, we consider the absorption and the reflectivity,
which are the optical quantities usually measured in real experimental
setups. Let us consider that the electric field of the probe pulse
is polarized along some specific direction determined by the unit
vector $\boldsymbol{u}_{\mathrm{pr}}$, and its reflection is measured
along the direction $\boldsymbol{u}_{\mathrm{pr}}^{\prime}$. Therefore,
the relevant element of our optical conductivity tensor would be $\boldsymbol{u}_{\mathrm{pr}}^{\prime}\cdot\boldsymbol{\sigma}\left(\omega,t_{\mathrm{pr}}\right)\cdot\boldsymbol{u}_{\mathrm{pr}}$.
We assume that the off-diagonal elements of the optical conductivity
tensor are negligible, i.e., the only relevant case to be studied
is $\boldsymbol{u}_{\mathrm{pr}}=\boldsymbol{u}_{\mathrm{pr}}^{\prime}$.
It is worth noting that if the spin inversion symmetry is not broken,
one can consider only one spin projection in the calculations and
just double the final result for the optical conductivity.

From the optical conductivity one can obtain the dielectric function,
$\epsilon\left(\omega,t_{\mathrm{pr}}\right)$, as (see App.~\ref{sec:out-of-eq-optics})
\begin{equation}
\epsilon\left(\omega,t_{\mathrm{pr}}\right)=1+\frac{\mathrm{i}}{\omega\epsilon_{0}}\sigma\left(\omega,t_{\mathrm{pr}}\right),\label{eq:eps_sig}
\end{equation}
where $\epsilon_{0}$ is the vacuum dielectric constant, $\sigma\left(\omega,t_{\mathrm{pr}}\right)=\boldsymbol{u}_{\mathrm{pr}}\cdot\boldsymbol{\sigma}\left(\omega,t_{\mathrm{pr}}\right)\cdot\boldsymbol{u}_{\mathrm{pr}}$,
and $\epsilon\left(\omega,t_{\mathrm{pr}}\right)=\boldsymbol{u}_{\mathrm{pr}}\cdot\boldsymbol{\epsilon}\left(\omega,t_{\mathrm{pr}}\right)\cdot\boldsymbol{u}_{\mathrm{pr}}$.

As discussed in App.~\ref{sec:out-of-eq-optics}, when the probe
pulse frequency is much larger than the pump pulse one, one can use
the well-known formula for equilibrium reflectivity also to obtain
the reflectivity out of equilibrium. Then, for a s-polarized probe
pulse at an incident angle $\theta$, with frequency $\omega$ and
center $t_{\mathrm{pr}}$, the transient reflectivity is given by
\begin{equation}
R_{\theta}\left(\omega,t_{\mathrm{pr}}\right)=\left|\frac{\cos\theta-\sqrt{\epsilon\left(\omega,t_{\mathrm{pr}}\right)-\mathrm{\sin}^{2}\theta}}{\cos\theta+\sqrt{\epsilon\left(\omega,t_{\mathrm{pr}}\right)-\mathrm{\sin}^{2}\theta}}\right|^{2}.\label{eq:Refl}
\end{equation}

To analyze the experimental results \citep{inzani2022field}, we are
interested in the transient relative differential reflectivity, $\delta_{r}R_{\theta}\left(\omega,t_{\mathrm{pr}}\right)$,
defined as
\begin{equation}
\delta_{r}R_{\theta}\left(\omega,t_{\mathrm{pr}}\right)=\frac{R_{\theta}\left(\omega,t_{\mathrm{pr}}\right)-R_{\theta}^{\mathrm{eq}}\left(\omega\right)}{R_{\theta}^{\mathrm{eq}}\left(\omega\right)},\label{eq:dR}
\end{equation}
where $R_{\theta}^{\mathrm{eq}}\left(\omega\right)=R_{\theta}\left(\omega,t_{\mathrm{pr}}\rightarrow t_{\mathrm{ini}}\right)$
is the equilibrium reflectivity computed using the optical conductivity
given in Eq.~\ref{eq:sigma-eq-ana}.

To get more insight into the transient reflectivity and its frequency
content, one may perform a Fourier transform with respect to the time
$t_{\mathrm{pr}}$ and obtain 
\begin{equation}
\delta_{r}R_{\theta}\left(\omega,\omega^{\prime}\right)=\int_{-\infty}^{+\infty}\mathrm{e}^{\mathrm{i}\omega t_{\mathrm{pr}}}\delta_{r}R_{\theta}\left(\omega,t_{\mathrm{pr}}\right)dt_{\mathrm{pr}}.\label{eq:FTdR}
\end{equation}

Another interesting optical property is the absorption coefficient,
which is given by

\begin{equation}
\alpha\left(\omega,t_{\mathrm{pr}}\right)=\frac{\omega}{n_{\mathrm{refr}}\left(\omega,t_{\mathrm{pr}}\right)c}\Im\left[\epsilon\left(\omega,t_{\mathrm{pr}}\right)\right],\label{eq:alpha}
\end{equation}
where $c$ is the speed of light in vacuum and $n_{\mathrm{refr}}\left(\omega,t_{\mathrm{pr}}\right)=\Re\left[\sqrt{\epsilon\left(\omega,t_{\mathrm{pr}}\right)}\right]$
is the real transient refractive index. In our numerical calculations,
in order to study the transient behavior of the system, we will investigate
the transient differential absorption coefficient,
\begin{equation}
\delta\alpha\left(\omega,t_{\mathrm{pr}}\right)=\alpha\left(\omega,t_{\mathrm{pr}}\right)-\alpha^{\mathrm{eq}}\left(\omega\right),\label{eq:dalpha}
\end{equation}
where $\alpha^{\mathrm{eq}}\left(\omega\right)$ is the absorption
coefficient at equilibrium.

\subsection{Equilibrium properties\protect\label{subsec:Equilibrium-properties}}

In equilibrium, we have $P_{\boldsymbol{k},nn^{\prime}}\left(t\right)=\mathrm{e}^{-\mathrm{i}\omega_{\boldsymbol{k}n}\left(t-t_{\mathrm{ini}}\right)}\delta_{nn^{\prime}}$.
Inserting this into Eqs.~\ref{eq:sigma-1-P-t} and \ref{eq:sigma-2-P-t},
and performing some straightforward calculations, one obtains

\begin{equation}
\boldsymbol{\sigma}^{\mathrm{eq}}\left(\omega\right)=\boldsymbol{\sigma}^{\mathrm{eq},\mathrm{main}}\left(\omega\right)+\boldsymbol{\sigma}_{1}^{\mathrm{eq},\mathrm{tail}}\left(\omega\right)+\boldsymbol{\sigma}_{2}^{\mathrm{eq},\mathrm{tail}}\left(\omega\right),\label{eq:sigma-eq-ana}
\end{equation}
where
\begin{multline}
\boldsymbol{\sigma}^{\mathrm{eq},\mathrm{main}}\left(\omega\right)=\frac{\mathrm{i}e^{2}}{\hbar\mathcal{V}}\sum_{\boldsymbol{k}}\sum_{nn^{\prime}}\\
\frac{\boldsymbol{J}_{\boldsymbol{k},n^{\prime}n}\boldsymbol{J}_{\boldsymbol{k},nn^{\prime}}\left(f_{\boldsymbol{k},n^{\prime}}-f_{\boldsymbol{k},n}\right)}{\omega_{\boldsymbol{k},nn^{\prime}}\left(\omega-\omega_{\boldsymbol{k},nn^{\prime}}+\mathrm{i}0^{+}\right)},\label{eq:sigma-eq-clean}
\end{multline}
and
\begin{multline}
\boldsymbol{\sigma}_{1}^{\mathrm{eq},\mathrm{tail}}\left(\omega\right)=-\frac{\mathrm{i}e^{2}}{\hbar^{3}\mathcal{V}}\sum_{\boldsymbol{k}}\sum_{nn^{\prime}}\\
\frac{\boldsymbol{\eta}_{\boldsymbol{k},n^{\prime}n}\boldsymbol{\eta}_{\boldsymbol{k},nn^{\prime}}\left(f_{\boldsymbol{k},n^{\prime}}-f_{\boldsymbol{k},n}\right)}{\omega_{\boldsymbol{k}nn^{\prime}}\left(\omega+\mathrm{i}0^{+}\right)},\label{eq:sigma-eq-tail-1}
\end{multline}
and
\begin{align}
 & \boldsymbol{\sigma}_{2}^{\mathrm{eq},\mathrm{tail}}\left(\omega\right)=\frac{\mathrm{i}e^{2}}{\hbar^{2}\mathcal{V}}\sum_{\boldsymbol{k}}\sum_{nn^{\prime}}\frac{f_{\boldsymbol{k},n}\boldsymbol{\xi}_{\boldsymbol{k},nn^{\prime}}\delta_{nn^{\prime}}}{\omega+\mathrm{i}0^{+}},\label{eq:sigma-eq-tail-2}
\end{align}
in which $\omega_{\boldsymbol{k},nn^{\prime}}=\omega_{\boldsymbol{k},n}-\omega_{\boldsymbol{k},n^{\prime}}=\left(\varepsilon_{\boldsymbol{k},n}-\varepsilon_{\boldsymbol{k},n^{\prime}}\right)/\hbar$,
and $\boldsymbol{J}_{\boldsymbol{k},nn^{\prime}}=\boldsymbol{J}_{\boldsymbol{k},nn^{\prime}}\left(t\rightarrow t_{\mathrm{ini}}\right)=\boldsymbol{\eta}_{\boldsymbol{k},nn^{\prime}}/\hbar+\mathrm{\mathrm{i}}\omega_{\boldsymbol{k},nn^{\prime}}\boldsymbol{D}_{\boldsymbol{k},nn^{\prime}}$.

$\boldsymbol{\sigma}^{\mathrm{eq},\mathrm{main}}\left(\omega\right)$
is the main contribution to the equilibrium conductivity, which doesn't
diverge in the limit $\omega\rightarrow0$ (see below). On the other
hand, $\boldsymbol{\sigma}^{\mathrm{eq},\mathrm{tail}}\left(\omega\right)=\boldsymbol{\sigma}_{1}^{\mathrm{eq},\mathrm{tail}}\left(\omega\right)+\boldsymbol{\sigma}_{2}^{\mathrm{eq},\mathrm{tail}}\left(\omega\right)$
in general diverges in the static limit, $\omega\rightarrow0$, which
gives the expected Drude behavior. However, for semiconductors at
zero temperature, one can show that $\boldsymbol{\sigma}_{1}^{\mathrm{eq},\mathrm{tail}}\left(\omega\right)=-\boldsymbol{\sigma}_{2}^{\mathrm{eq},\mathrm{tail}}\left(\omega\right)$
and, therefore, $\boldsymbol{\sigma}^{\mathrm{eq},\mathrm{tail}}\left(\omega\right)$
identically vanishes (see App.~\ref{sec:Cancellation-of-tails}).
In numerical calculations for \emph{real} materials, the relative
difference of $\boldsymbol{\sigma}_{1}^{\mathrm{eq},\mathrm{tail}}\left(\omega\right)$
and $-\boldsymbol{\sigma}_{2}^{\mathrm{eq},\mathrm{tail}}\left(\omega\right)$
can be used to check if the adopted \textbf{k} grid is dense enough
and/or if the number of energy bands taken into account is sufficient.

After this discussion on the optical conductivity, it is relevant
to study the behavior of the equilibrium dielectric function in the
static limit, $\omega\rightarrow0$. For the sake of simplicity, we
consider zero temperature, so that the tails in the optical conductivity
cancel each other and we get $\sigma^{\mathrm{eq}}\left(\omega\right)\rightarrow\sigma^{\mathrm{eq},\mathrm{main}}\left(\omega\right)$,
where $\sigma^{\mathrm{eq}}\left(\omega\right)=\boldsymbol{u}_{\mathrm{pr}}\cdot\boldsymbol{\sigma}^{\mathrm{eq}}\left(\omega\right)\cdot\boldsymbol{u}_{\mathrm{pr}}$
. After some tricky calculations, one can show that the static limit
of the equilibrium dielectric function is given by

\begin{equation}
\epsilon^{\mathrm{eq}}\left(\omega\rightarrow0\right)=1+\left(1+\mathrm{i}\frac{0^{+}}{\omega}\right)C,\label{eq:eps_w0}
\end{equation}
where 
\begin{equation}
C=\frac{e^{2}}{\hbar\epsilon_{0}\mathcal{V}}\sum_{\boldsymbol{k}}\sum_{nn^{\prime}}\frac{\left|J_{\boldsymbol{k},n^{\prime}n}\right|^{2}}{\omega_{\boldsymbol{k},nn^{\prime}}^{3}}\left(f_{\boldsymbol{k},n^{\prime}}-f_{\boldsymbol{k},n}\right),
\end{equation}
in which $J_{\boldsymbol{k},n^{\prime}n}=\boldsymbol{u}_{\mathrm{pr}}\cdot\boldsymbol{J}_{\boldsymbol{k},n^{\prime}n}$.
$C$ is in general a real positive number, so that the static dielectric
function is larger than 1, as it is quite well known for semiconductors.
The imaginary part of $\epsilon^{\mathrm{eq}}\left(\omega\rightarrow0\right)$
is given by $\frac{0^{+}}{\omega}C$, which vanishes if one takes
the correct order of the limits: first $0^{+}\rightarrow0$ and then
$\omega\rightarrow0$. In numerical calculations, at low frequencies,
$0^{+}$ is a finite number and $\frac{0^{+}}{\omega}$ cannot vanish
in the limit of $\omega\rightarrow0$. Actually, our results are valid
only if the probe frequency is much larger than $0^{+}$. As we will
describe below, in order to have a smaller $0^{+}$, one needs a higher
energy resolution and consequently a finer grid in \textbf{k} space.
This is consistent with the fact that, to study the behavior of a
system in some energy regime, it is needed a sufficiently fine energy
resolution of the energy spectrum of the system.

Having computed the optical conductivity, one can obtain the dielectric
function in equilibrium and in turn, the real refractive index in
equilibrium. For the frequencies where the real equilibrium refractive
index, $n_{\mathrm{refr}}^{\mathrm{eq}}\left(\omega\right)$, is smaller
than 1, one can define the critical incident angle, $\theta_{C}\left(\omega\right)$,
as
\begin{equation}
\theta_{C}\left(\omega\right)=\sin^{-1}\left[n_{\mathrm{refr}}^{\mathrm{eq}}\left(\omega\right)\right].
\end{equation}
In absence of absorption, for incident angles above $\theta_{C}\left(\omega\right)$,
we have total external reflection, i.e., the reflectivity is equal
to 1. In presence of absorption, the reflectivity is smaller than
1 even for incident angles above $\theta_{C}\left(\omega\right)$.
On the other hand, for $n_{\mathrm{refr}}^{\mathrm{eq}}\left(\omega\right)\geq1$
we have $\theta_{C}\left(\omega\right)=90{^\circ}$.

\subsection{Guidelines for numerical calculations\protect\label{subsec:Guidelines-for-numerical}}

In this section, we give some guidelines that can help to perform
the numerical calculations of the reflectivity and of the absorption
in different pump-probe setups.

\subsubsection{Rewriting the expression for the optical conductivity to simplify
its numerical calculation\protect\label{subsec:Rewriting-sigma-xyz}}

At a first glance, it may seem that Eq.~\ref{eq:sigma-1-P-t} is
computationally very expensive. It requires summing over many indices
and this can be quite heavy for \emph{real} materials. Moreover, one
has to calculate a time integral for each pair of $t$ and $t_{\mathrm{pr}}$.
However, with some careful manipulations, these complexities can be
simplified. One can rewrite Eq.~\ref{eq:sigma-1-P-t} as follows

\begin{multline}
\boldsymbol{\sigma}_{1}\left(t,t_{\mathrm{pr}}\right)=\frac{\mathrm{i}e}{\hbar\mathcal{V}}\theta\left(t-t_{\mathrm{pr}}\right)\\
\sum_{\boldsymbol{k}}\mathrm{Tr}\left\{ Z_{\boldsymbol{k}}\left(t\right)\circ\left[Y_{\boldsymbol{k}}\left(t\right)-Y_{\boldsymbol{k}}\left(t_{\mathrm{pr}}\right)+X_{\boldsymbol{k}}\left(t_{\mathrm{pr}}\right)\right]\right\} ,\label{eq:sig_ZYX}
\end{multline}
where $\circ$ stands for the matrix product in the band space and
$Z_{\boldsymbol{k}}\left(t\right)$, $Y_{\boldsymbol{k}}\left(t\right)$
and $X_{\boldsymbol{k}}\left(t\right)$ are matrices with the following
elements:
\begin{multline}
Z_{\boldsymbol{k},n_{1}^{\prime}n_{2}^{\prime}}\left(t\right)=\\
\sum_{n_{1}n_{2}}\boldsymbol{J}_{\boldsymbol{k},n_{1}n_{2}}\left(t\right)P_{\boldsymbol{k},n_{1}n_{1}^{\prime}}^{\star}\left(t\right)P_{\boldsymbol{k},n_{2}n_{2}^{\prime}}\left(t\right)\left(f_{\boldsymbol{k}n_{1}^{\prime}}-f_{\boldsymbol{k}n_{2}^{\prime}}\right),
\end{multline}
\begin{multline}
Y_{\boldsymbol{k},n_{2}^{\prime}n_{1}^{\prime}}\left(t\right)=-\frac{e}{\hbar}\sum_{n_{3}n_{4}}\int_{t_{\mathrm{ini}}}^{t}dt^{\prime}\\
\boldsymbol{V}_{\boldsymbol{k},n_{3}n_{4}}\left(t^{\prime}\right)P_{\boldsymbol{k},n_{3}n_{2}^{\prime}}^{\star}\left(t^{\prime}\right)P_{\boldsymbol{k},n_{4}n_{1}^{\prime}}\left(t^{\prime}\right),\label{eq:Y}
\end{multline}
\begin{align}
X_{\boldsymbol{k},n_{2}^{\prime}n_{1}^{\prime}}\left(t\right) & =e\sum_{n_{3}n_{4}}\boldsymbol{D}_{\boldsymbol{k},n_{3}n_{4}}\left(t\right)P_{\boldsymbol{k},n_{3}n_{2}^{\prime}}^{\star}\left(t\right)P_{\boldsymbol{k},n_{4}n_{1}^{\prime}}\left(t\right).
\end{align}
Using such matrices, the numerical calculation of $\boldsymbol{\sigma}_{1}\left(t,t_{\mathrm{pr}}\right)$
becomes very fast and efficient as it avoids summing over several
indices, which would be computationally very expensive in simulating
\emph{real} materials where one deals with many bands.

It is worth noting that, in the numerical calculations, where it is
obviously needed to discretize the $t$ time axis, at each step of
time, $t_{m}$, one can compute $Y_{\boldsymbol{k},n_{2}^{\prime}n_{1}^{\prime}}\left(t_{m}\right)$
(Eq.~\ref{eq:Y}) recursively from $Y_{\boldsymbol{k},n_{2}^{\prime}n_{1}^{\prime}}\left(t_{m-1}\right)$
and approximating the remaining integral between $t_{m-1}$ and $t_{m}$.
Therefore, there is no need to perform the increasingly expensive
integration between $t_{\mathrm{ini}}$ and $t_{m}$ for each $t_{m}$.

\subsubsection{Finite numerical momentum grid and corresponding damping factor\protect\label{subsec:Finite-k-grid}}

For the numerical calculations, we need to consider a finite grid
in the first Brillouin zone to sample the momentum $\boldsymbol{k}$.
Such a grid will have a total number of $\boldsymbol{k}$ points $N_{\mathrm{grid}}$
and all summations over $\boldsymbol{k}$ should be performed as $\frac{1}{\mathcal{V}}\sum_{\boldsymbol{k}}\rightarrow\frac{1}{v_{\mathrm{uc}}N_{\mathrm{grid}}}\sum_{\boldsymbol{k}\in\mathrm{grid}}$
where $v_{\mathrm{uc}}$ is the unit cell volume. Another issue which
is directly connected to the numerical $\boldsymbol{k}$ grid is the
finite damping factor, or equivalently, level broadening, $0^{+}$
. In real experimental setups, one always deals with a finite level
broadening originating from several scattering mechanisms (such as
electron-phonon interaction, disorder, etc.) present in the system,
or even from the measurement procedure. Such a level broadening brings
a finite energy resolution, which justifies the use of a finite momentum
grid instead of the almost continuous \emph{real} one. The sparser
is the $\boldsymbol{k}$ grid, implying a lower and lower achievable
energy resolution, the larger has to be the value of $0^{+}$. This
is crucial in the numerical calculations: if the value of $0^{+}$
is smaller than the band energy resolution provided by the momentum
grid, the features corresponding to individual $\boldsymbol{k}$ points
show up in the final results making them artificially spiky. Then,
if one is forced, by the available computational resources, to use
a sparser $\boldsymbol{k}$ grid than the one consistent with the
actual damping in the \emph{real} material under analysis (as it is
usually the case), one has to use a larger value of $0^{+}$ than
the physical one and this comes at the cost of suppressing possibly
relevant physical features (see App.~\ref{sec:zp_study}).

\subsubsection{Analytical simplification regarding the time range after the application
of the pump pulse\protect\label{subsec:Analytical-after-pump}}

The time step for the Fourier transformation of the optical conductivity,
Eq.~\ref{eq:FT-sigma}, should be small enough so that its reciprocal
is larger than the frequency of the probe pulse. On the other hand,
the values of $t_{\mathrm{pr}}$ suffer no physical or mathematical
restrictions, and one can freely choose the instants of time at which
to probe the system. The initial time in the numerical calculations,
$t_{\mathrm{ini}}\rightarrow-\infty$, is simple to choose: it should
be a time where the pump pulse is negligible, and also should meet
the criteria $t_{\mathrm{ini}}<t_{\mathrm{pr}}$.

However, for the final time, $t_{\mathrm{fin}}\rightarrow+\infty$,
in addition to the condition that the pump pulse should become negligible
after its application and that $t_{\mathrm{fin}}>t_{\mathrm{pr}}$
for every $t_{\mathrm{pr}}$, it is needed that the integrand of Eq.~\ref{eq:FT-sigma}
becomes negligible at $t_{\mathrm{fin}}$. That is, for every $t_{\mathrm{pr}}$,
it is required that $\mathrm{e}^{-0^{+}\left(t_{\mathrm{fin}}-t_{\mathrm{pr}}\right)}\simeq0$.
One can always choose a large enough value for $t_{\mathrm{fin}}$
and solve this problem. However, in order to increase the speed of
computations, one can choose $t_{\mathrm{fin}}$ to meet only one
criteria, i.e., the vanishing of pump and probe pulses, and write

\begin{align}
\boldsymbol{\sigma}\left(\omega,t_{\mathrm{pr}}\right) & =\bar{\boldsymbol{\sigma}}\left(\omega,t_{\mathrm{pr}}\right)+\boldsymbol{\sigma}^{\mathrm{a.p.}}\left(\omega,t_{\mathrm{pr}}\right),
\end{align}
where the first term on the right hand side is the result of the numerical
calculations
\begin{equation}
\bar{\boldsymbol{\sigma}}\left(\omega,t_{\mathrm{pr}}\right)=\int_{t_{\mathrm{pr}}}^{t_{\mathrm{fin}}}dt\mathrm{e}^{\mathrm{i}\left(\omega+\mathrm{i}0^{+}\right)\left(t-t_{\mathrm{pr}}\right)}\boldsymbol{\sigma}\left(t,t_{\mathrm{pr}}\right),
\end{equation}
while the second term, which we dub \emph{after-pump} contribution,
is given by
\begin{equation}
\boldsymbol{\sigma}^{\mathrm{a.p.}}\left(\omega,t_{\mathrm{pr}}\right)=\int_{t_{\mathrm{fin}}}^{\infty}dt\mathrm{e}^{\mathrm{i}\left(\omega+\mathrm{i}0^{+}\right)\left(t-t_{\mathrm{pr}}\right)}\boldsymbol{\sigma}\left(t,t_{\mathrm{pr}}\right),
\end{equation}
and can be calculated analytically, as we are going to describe in
the following. For $t>t_{\mathrm{fin}}$, the pump pulse becomes negligible,
and the projection coefficients follow a trivial dynamics:
\begin{equation}
P_{\boldsymbol{k}nn^{\prime}}\left(t\right)=\mathrm{e}^{-\mathrm{i}\omega_{\boldsymbol{k}n}\left(t-t_{\mathrm{fin}}\right)}P_{\boldsymbol{k}nn^{\prime}}\left(t_{\mathrm{fin}}\right),\qquad t\geq t_{\mathrm{fin}},
\end{equation}
and the pump-pulse dependent observable matrix elements (such as velocities,
dipole elements, currents, etc.) return to their equilibrium values.
After some lengthy but straightforward calculations one can show that
\begin{multline}
\boldsymbol{\sigma}_{1}^{\mathrm{a.p.}}\left(\omega,t_{\mathrm{pr}}\right)=-\frac{\mathrm{i}e}{\hbar\mathcal{V}}\sum_{\boldsymbol{k}}\left\{ W_{\boldsymbol{k}}\left(\omega,t_{\mathrm{fin}},t_{\mathrm{pr}}\right)\right.\\
\left.+\mathrm{Tr}\left[Q_{\boldsymbol{k}}\left(\omega,t_{\mathrm{fin}},t_{\mathrm{pr}}\right)\circ S_{\boldsymbol{k}}\left(t_{\mathrm{fin}},t_{\mathrm{pr}}\right)\right]\right\} ,
\end{multline}
where
\begin{multline}
Q_{\boldsymbol{k},n^{\prime}n}\left(\omega,t_{\mathrm{fin}},t_{\mathrm{pr}}\right)=\sum_{n_{1}n_{2}}\\
P_{\boldsymbol{k},n_{2}n^{\prime}}^{*}\left(t_{\mathrm{fin}}\right)\boldsymbol{J}_{\boldsymbol{k},n_{2}n_{1}}P_{\boldsymbol{k},n_{1}n}\left(t_{\mathrm{fin}}\right)\\
\frac{\mathrm{i}\mathrm{e}^{\left(\mathrm{i}\omega-0^{+}\right)\left(t_{\mathrm{fin}}-t_{\mathrm{pr}}\right)}}{\left(\omega-\omega_{\boldsymbol{k},n_{1}n_{2}}+\mathrm{i}0^{+}\right)}\left(f_{\boldsymbol{k}n^{\prime}}-f_{\boldsymbol{k}n}\right),
\end{multline}
\begin{equation}
S_{\boldsymbol{k}}\left(t_{\mathrm{fin}},t_{\mathrm{pr}}\right)=Y_{\boldsymbol{k}}\left(t_{\mathrm{fin}}\right)-Y_{\boldsymbol{k}}\left(t_{\mathrm{pr}}\right)-X_{\boldsymbol{k}}\left(t_{\mathrm{pr}}\right),
\end{equation}
\begin{multline}
W_{\boldsymbol{k}}\left(\omega,t_{\mathrm{fin}},t_{\mathrm{pr}}\right)=-\frac{e}{\hbar}\sum_{n_{1}n_{2}}\\
\frac{\mathrm{e}^{\left(\mathrm{i}\omega-0^{+}\right)\left(t_{\mathrm{fin}}-t_{\mathrm{pr}}\right)}\boldsymbol{J}_{\boldsymbol{k},n_{2}n_{1}}}{\omega-\omega_{\boldsymbol{k},n_{1}n_{2}}+\mathrm{i}0^{+}}\\
\sum_{n}\left(\frac{\boldsymbol{\eta}_{\boldsymbol{k},n_{1}n}}{\omega-\omega_{\boldsymbol{k},nn_{2}}+\mathrm{i}0^{+}}N_{\boldsymbol{k},n_{2}n}\left(t_{\mathrm{fin}}\right)\right.\\
\left.-\frac{\boldsymbol{\eta}_{\boldsymbol{k},nn_{2}}}{\omega-\omega_{\boldsymbol{k},n_{1}n}+\mathrm{i}0^{+}}N_{\boldsymbol{k},nn_{1}}\left(t_{\mathrm{fin}}\right)\right),
\end{multline}
and
\begin{equation}
\boldsymbol{\sigma}_{2}^{\mathrm{a.p.}}\left(\omega,t_{\mathrm{pr}}\right)=\frac{e}{\mathcal{V}}\sum_{\boldsymbol{k}}\sum_{n_{1}n_{2}}\frac{\mathrm{i}N_{\boldsymbol{k},n_{1}n_{2}}\left(t_{\mathrm{fin}}\right)\frac{\delta\boldsymbol{J}_{\boldsymbol{k}.n_{1}n_{2}}}{\delta\boldsymbol{A}}}{\omega_{\boldsymbol{k},n_{1}n_{2}}+\omega+\mathrm{i}0^{+}},
\end{equation}
and clearly, we have 
\begin{equation}
\boldsymbol{\sigma}^{\mathrm{a.p.}}\left(\omega,t_{\mathrm{pr}}\right)=\boldsymbol{\sigma}_{1}^{\mathrm{a.p.}}\left(\omega,t_{\mathrm{pr}}\right)+\boldsymbol{\sigma}_{2}^{\mathrm{a.p.}}\left(\omega,t_{\mathrm{pr}}\right).
\end{equation}

\subsubsection{Approximations considering the core levels\protect\label{subsec:Approximations-core-levels}}

Another numerical efficiency improvement can be achieved by separating
the \emph{core} bands, which are far below the Fermi energy (i.e.,
the energy gaps between them and the other bands are out of resonance
by far with respect to the pump photon energy), from the valence and
the conduction bands. Obviously, the core bands are completely filled
at equilibrium, i.e., $f_{\boldsymbol{k}n_{\mathrm{core}}}=1$, and
we can consider negligible the effect of the pump pulse on them. Therefore,
one can write
\begin{equation}
P_{\boldsymbol{k}n_{\mathrm{core}}n}\left(t\right)=P_{\boldsymbol{k}nn_{\mathrm{core}}}\left(t\right)=\mathrm{e}^{-\mathrm{i}\omega_{\boldsymbol{k}n_{\mathrm{core}}}\left(t-t_{\mathrm{ini}}\right)}\delta_{nn_{\mathrm{core}}},
\end{equation}
where $\omega_{\boldsymbol{k},n_{\mathrm{core}}}=\varepsilon_{\boldsymbol{k},n_{\mathrm{core}}}/\hbar$.

Another approximation regarding the \emph{core} bands is to consider
only the energy gaps between them and the valence and conduction bands
in computing the optical conductivity, and neglect the contributions
coming from the energy gaps between valence and conduction bands.
For more details about this approximation see App.~\ref{sec:core-surface-cond}.

\subsubsection{Quasi-static approximation\protect\label{subsec:QS-main-text}}

Finally, a possible approximation one may think of is to assume that
on the time scale corresponding to the probe frequencies, the pump-induced
evolution can be considered quasi-static. Even though such an approximation
speeds up the calculations, as it allows for a significant part of
the job to be done analytically, it cannot reproduce some relevant
features of the actual behavior of the optical properties. For further
discussions about this approximation see App.~\ref{sec:Quasi-static-approximation}.

\section{Results\protect\label{sec:Two-band_system}}

\begin{figure*}[t]
\centering{}\includegraphics[width=18cm]{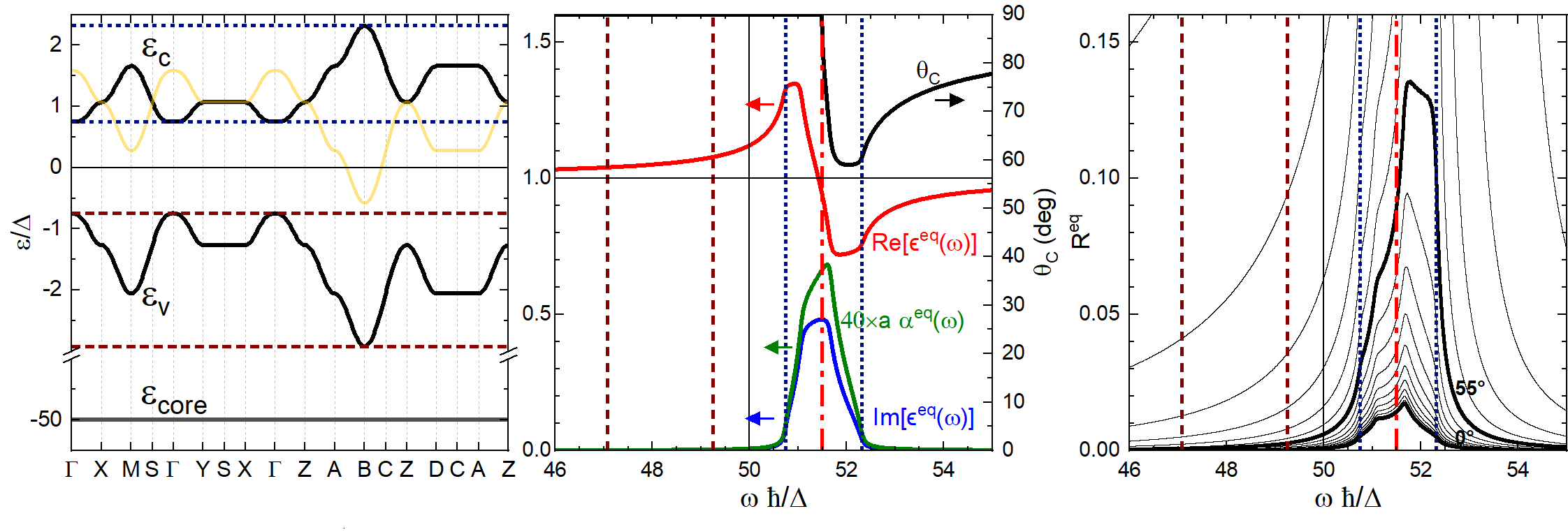}\caption{(left) Equilibrium bands along the main path passing through the high-symmetry
points in the \textbf{k} space of the cubic lattice. The one-photon
side-band with photon energy $2.33\Delta$ with respect to the VB
is also shown (solid yellow). At the \textbf{k} points where this
side-band crosses the CB, we have one-photon resonances and Rabi-like
excitations. (middle) The critical angle, the real and imaginary parts
of the dielectric function, and the absorption coefficient, at equilibrium,
as functions of the probe photon energy $\omega$. The absorption
coefficient has been multiplied by the lattice constant $a$ to make
it dimensionless and by 40 to make it visible on the range of the
left scale. A dashed-dot vertical red line marks the frequency below
which the critical angle is just 90°. (right) The equilibrium reflectivity
at different incident angles, starting from 0° (normal incidence)
and increasing with the steps of 5°, vs the probe photon energy $\omega$.
The top and the bottom of the VB (CB) are marked by dashed-wine (dotted-blue)
lines in all panels.\protect\label{fig:Equilibrium}}
\end{figure*}

\subsection{The system\protect\label{subsec:The-system}}

In this section, we consider the minimal model for a semiconducting
material: a cubic lattice with lattice constant $a$ and one valence
(VB), one conduction (CB), and one \emph{core} band. Such a model
allows us to determine what kinds of phenomena can, in principle,
emerge in \emph{real} experimental setups without being lost in the
complications arising from the specific features of a particular \emph{real}
material, even though the theory does not have any limitation in the
lattice structure or the number of bands that can be studied. We consider
the VB and the CB to derive from two Wannier states with onsite energies
$\tilde{T}_{\boldsymbol{0},1,1}=-1.65\Delta$ and $\tilde{T}_{\boldsymbol{0},2,2}=1.35\Delta$,
diagonal first-neighbor hoppings $\tilde{T}_{\mathbf{a},1,1}=0.2\Delta$
and $\tilde{T}_{\mathbf{a},2,2}=-0.15\Delta$, and off-diagonal first-neighbor
hoppings $\tilde{T}_{\mathbf{a},1,2}=\tilde{T}_{\mathbf{a},2,1}=-0.1\Delta$,
where, as mentioned before, $\tilde{T}_{\boldsymbol{R},\nu,\nu^{\prime}}$
is the hopping matrix between two Wannier states $\nu$ and $\nu^{\prime}$
centered at distance $\boldsymbol{R}$, and $\mathbf{a}\in\left\{ a\left(\pm1,0,0\right),a\left(0,\pm1,0\right),a\left(0,0,\pm1\right)\right\} $.
$\Delta$ is the unit of energy and can be adjusted to obtain the
desired band-gap energies. As a reference, $\Delta=\unit[0.5]{eV}$
gives a band gap of $\unit[0.75]{eV}$ at $\Gamma$. The \emph{core}
band is assumed to be flat, with an energy $\varepsilon_{\mathrm{core}}=-50\Delta$.
To focus on the pump-pulse-induced dynamics, without the need to disentangle
it from too specific details of the probe-pulse coupling to the system,
we consider a local (momentum-independent) dipole matrix element $\boldsymbol{D}=\mathrm{i}0.05a\hat{\mathbf{j}}$
between the \emph{core} level and both the VB and the CB. We sample
the \textbf{k} space by a cubic grid of size $32\times32\times32$
pinned at $\Gamma$. The damping factor for computing the optical
conductivity in the XUV regime is chosen to be $0^{+}=0.05\Delta/\hbar$
(see App.~\ref{fig:zp_stud}).

In Fig.~\ref{fig:Equilibrium} (left panel), we show the band energies
along the main path passing through the high-symmetry points in the
\textbf{k} space of the cubic lattice~\citep{eskandari2023dynamical}
as well as the one-photon sideband of the VB corresponding to a photon
energy $2.33\Delta$. Pumping the system with such a frequency, at
the \textbf{k}-points where this sideband crosses the CB, we have
the exact one-photon resonances and, therefore, Rabi-like excitations.
Given a \textbf{k}-point, pumping fields with different polarization
directions couple very differently to the system because the velocity
$\boldsymbol{\eta}_{\boldsymbol{k}}$ is direction-dependent in \textbf{k}
space.

In the middle panel of Fig.~\ref{fig:Equilibrium}, we plot the critical
angle, the real and imaginary parts of the dielectric function, and
the absorption coefficient, in equilibrium, as functions of the probe
photon energy, $\omega$. At the moment, we are interested in the
XUV regime, which we define as the range of the energy gaps between
the \emph{core} band and the VB/CB: this requires that the \emph{real}
material has at least one \emph{core} band in the XUV energy range
below the Fermi energy. According to this, the probe explores the
available states in the VB and the CB by exciting electrons from the
\emph{core} band to such bands. Consequently, given that our \emph{core}
band is \emph{flat}, every probe photon energy $\hbar\omega$ corresponds
to the energy $\hbar\omega-\left|\varepsilon_{\mathrm{core}}\right|$
in the VB/CB energies. This is implicitly understood hereafter, and
every optical feature at a given XUV photon energy, $\hbar\omega$,
is possibly connected to the related band energy $\varepsilon\left(\boldsymbol{k}\right)$
after subtracting the \emph{core} energy, $\left|\varepsilon_{\mathrm{core}}\right|$.
Accordingly, the edges of the VB (CB) are marked by dashed-wine (dotted-blue)
vertical lines in all panels of Fig.~\ref{fig:Equilibrium} at the
\emph{proper} XUV photon energy, panel by panel.

$\Im\left[\epsilon^{\mathrm{eq}}\left(\omega\right)\right]$ has a
behavior very similar to the absorption coefficient, $\alpha^{\mathrm{eq}}\left(\omega\right)$,
and both of them are finite only in the CB energy range. This can
be understood by noting that in equilibrium, the XUV photons can be
absorbed only by exciting the electrons from the \emph{core} band
to the empty states of CB. However, the finite numerical value of
$0^{+}$ gives a small broadening in $\Im\left[\epsilon^{\mathrm{eq}}\left(\omega\right)\right]$
and $\alpha^{\mathrm{eq}}\left(\omega\right)$ outside of the energy
range of the CB. One should also consider that in real systems $0^{+}$
is finite, as several kinds of imperfections and interactions/couplings
make the lifetime of the excited states finite and broaden the energy
bands.

The real part of the dielectric function, $\Re\left[\epsilon^{\mathrm{eq}}\left(\omega\right)\right]$,
collects contributions from each \textbf{k-}point (Eq.~\ref{eq:sigma-eq-clean})
in the form of a Cauchy\textquoteright s principal part, centered
at its corresponding CB energy, and therefore, outside of the energy
range of the CB, $\Re\left[\epsilon^{\mathrm{eq}}\left(\omega\right)\right]$
does not vanish, and instead, has a hyperbolic-like drop vs energy
with a tail that is finite even in the energy range of the VB. Even
though the VB does not contribute to the optical properties at equilibrium,
this tail results in a non-zero equilibrium reflectivity at the VB
energy range, as discussed in the following. Generally, the non-locality
in energy (long tails) of Cauchy\textquoteright s principal part makes
some of the optical properties at each photon energy not only affected
by the band structure at such energy but also by the overall band
structure.

In the middle panel of Fig.~\ref{fig:Equilibrium}, we also have
the critical angle, $\theta_{C}$. Here, we assumed the reflection
of an s-polarized probe with an incident angle measured with respect
to the normal to the sample interface. The vertical red dash-dot line
shows the probe photon frequency where $\Re\left[n_{\mathrm{refr}}^{\mathrm{eq}}\right]=1$,
so that above (below) this frequency, $\theta_{C}$ is less than (equal
to) 90°. In the experimental setups, the incident angle of the probe
is chosen to be just below the critical angle in the range of probe
photon energies. This guarantees a high signal-to-noise-ratio in the
reflected beam and makes the results more precise and reliable. According
to this, in our case, we will choose the incident angle to be 55°.

In Fig.~\ref{fig:Equilibrium} (right panel), we show the equilibrium
reflectivities vs the probe photon energy, $\omega$, for different
incident angles starting from 0° (normal incidence) and increasing
in steps of 5°. The reflectivity eventually tends to zero for the
energies well outside of the CB range, even though it's not localized
in energy to the CB, because of the non-local-in-energy character
of $\Re\left[\epsilon^{\mathrm{eq}}\left(\omega\right)\right]$, as
explained above.

We apply to the system a pump pulse of the form $\boldsymbol{A}_{\mathrm{pu}}\left(t\right)=A_{\mathrm{pu}}\left(t\right)\hat{\mathbf{j}}$
where $A\left(t\right)$ is given by

\begin{equation}
A_{\mathrm{pu}}\left(t\right)=A_{0}e^{-\left(4\ln2\right)t^{2}/\tau_{\mathrm{pu}}^{2}}\cos\left(\omega_{\mathrm{pu}}t\right),\label{eq:A_t}
\end{equation}
in which, if not otherwise stated, the center of the pump pulse is
taken as the origin of time, the FWHM of the pump pulse is $\tau_{\mathrm{pu}}=7\hbar/\Delta$
, the frequency of the pump is $\omega_{\mathrm{pu}}=2.33\Delta/\hbar$
, and the pump amplitude is $A_{0}=0.4\pi\hbar/ae$. The probe has
the same polarization of the pump: $\boldsymbol{u}_{\mathrm{pr}}=\hat{\mathbf{j}}$.

\begin{figure*}[t]
\centering{}%
\begin{tabular}{c}
\includegraphics[width=19cm]{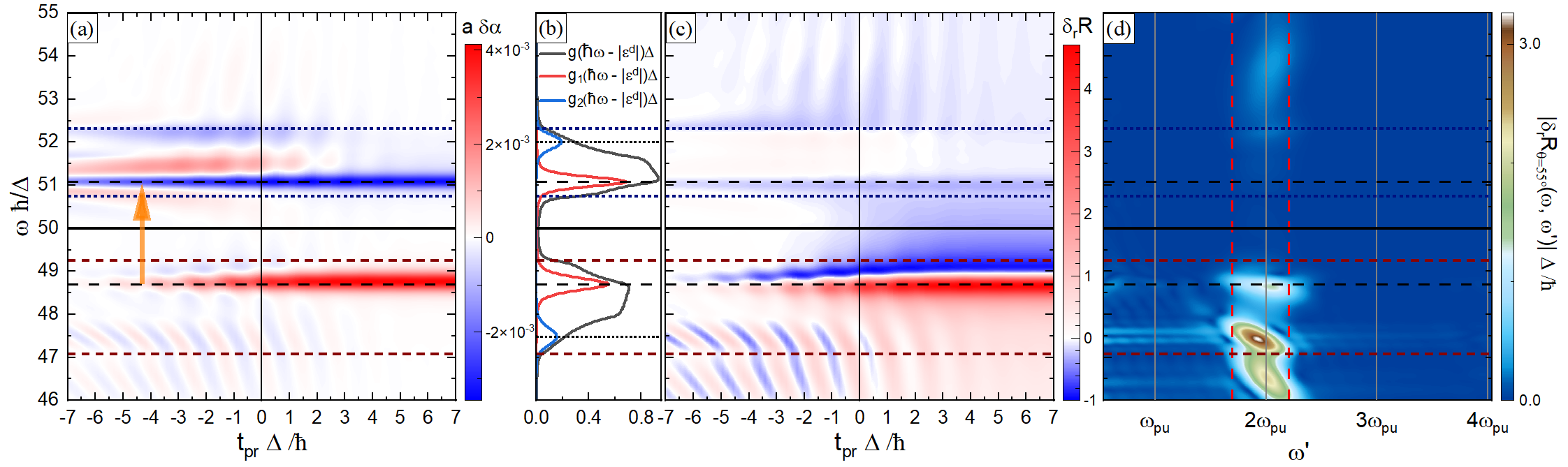}\tabularnewline
\end{tabular}\caption{Transient optical behavior of the pumped system for the case of Peierls
substitution coupling. (a) The transient differential absorption coefficient,
as given by Eq.~\ref{eq:dalpha}, vs probe time, $t_{\mathrm{pr}}$,
and probe photon frequency, $\omega$. The orange arrow determines
the main electronic excitation induced by the pump, which is the one-photon
resonance. (b) DOS and the one- and two-photon DORS vs energy difference
to the \emph{core} band. (c) Transient relative differential reflectivity
at the incident angle of $\theta=55{^\circ}$, as defined in Eq.~\ref{eq:dR},
vs delay time and probe photon energy. (d) The amplitude of the Fourier
transformation of the transient reflectivity, $\left|\delta R_{\theta=55{^\circ}}\left(\omega,\omega^{\prime}\right)\right|$,
see Eq.~\ref{eq:FTdR} (the low frequency region, $\omega^{\prime}<0.2\Delta/\hbar$,
is not shown). The horizontal black-dashed lines indicate the peaks
of the one-photon DORS, and the edges of VB (CB) are shown with dashed-wine
(dotted-blue) lines. \protect\label{fig:dR55_D=00003D0}}
\end{figure*}

\subsection{Peierls substitution coupling\protect\label{subsec:Peierls-substitution-coupling}}

At first, we consider (i) the system coupled to the pump pulse via
the Peierls substitution and (ii) no dipole coupling between the VB
and the CB. In Fig.~\ref{fig:dR55_D=00003D0}, we present the transient
behavior of the pumped system in such a case. Fig.~\ref{fig:dR55_D=00003D0}(a)
shows the transient differential absorption coefficient, as given
by Eq.~\ref{eq:dalpha}. The main red stripe in the VB and blue stripe
in the CB coincide with the black-dashed lines that indicate the positions
of the main peaks of the density of one-photon resonant states in
each band, which is reported in panel (b). One should notice that
at the resonances, the hole and electron populations mainly follow
a Rabi-like behavior with a period much longer than the FWHM of the
pump pulse, which results in the \emph{accumulation} of the excitation
populations. The electrons pumped into the CB leave less available
states for the electrons to be excited by the probe pulse from the
\emph{core} band to the CB. Consequently, the absorption of the probe
pulse is reduced that results in a blue stripe at the resonant energies
of the CB. On the other hand, the photo-injected holes into the VB
make it possible for the probe to excite electrons into the VB, which
leads to a finite absorption and, hence, to the red stripe in the
resonant energies of the VB. As we discussed in Ref.~\citep{eskandari2023dynamical},
the pump pulse is not single frequency, and, therefore, the resonance
is not sharp and has a finite width, which explains the broadening
of the main red and blue stripes around the resonance energies. Out
of the resonance, the excitation population has some transient oscillations~\citep{eskandari2023dynamical}
with different amplitudes. The collective effects of all such oscillations
in energy and time (as well as of the resonant ones) yield the other
blue and red areas, as shown in the map, together with the modifications
induced by the pump in the band structure of the system, to be intended
as the time-dependent Hamiltonian matrix elements.

In Fig.~\ref{fig:dR55_D=00003D0}(b), we show the density of states
(DOS) and the one- and two-photon density of resonant states (DORS)
vs energy, as given in the App.~\ref{sec:Density-of-states}. The
peaks of the $n_{\mathrm{ph}}$-photon DORS determine the energies
at which one would expect the effects of resonances. Indeed, as we
already discussed, the blue and red stripes in the map of $\delta\alpha$,
in the CB and VB ranges, respectively, stem from the one-photon resonances.
Notably, the two-photon resonances are very weak in this specific
case and do not have any relevant effect on the scales of Fig.~\ref{fig:dR55_D=00003D0}.
This is mainly due to the peaks of the two-photon DORS being near
the edges of the bands, where we have very small velocities and, hence,
weak couplings to the pump pulse. This occurrence, together with the
two-photon resonances being of second order and having low densities,
results in a negligible role in the transient optical properties in
this case.

In Fig.~\ref{fig:dR55_D=00003D0}(c), we show the transient relative
differential reflectivity at the incident angle $\theta=55{^\circ}$,
as defined in Eq.~\ref{eq:dR}. In this map, we see that the peaks
of the one-photon DORS (indicated by black-dashed lines) almost coincide
with the narrow white regions, where $\delta_{r}R$ approaches zero.
If one does not consider the imaginary part of the refraction index
in the reflectivity calculation, they will coincide (not shown). This
shows that the white narrow regions in the map of $\delta_{r}R$ indicate
the absorption edges. Moreover, on this map, one can see the usual
fishbone structure.

In Fig.~\ref{fig:dR55_D=00003D0}(d), we show the amplitude of the
Fourier transformation of the transient reflectivity, $\left|\delta R_{\theta=55{^\circ}}\left(\omega,\omega^{\prime}\right)\right|$,
as defined in Eq.~\ref{eq:FTdR}. As it is clear from this map, the
oscillatory part is around twice the pump frequency, $\omega^{\prime}=2\omega_{\mathrm{pu}}$,
and in particular, we don't have any odd-pump-frequency component.
The absence of the odd-pump-frequency components in the system's response
can be explained as follows. Writing the optical conductivity as the
sum of the contributions from individual \textbf{k} points, i.e.,
$\boldsymbol{\sigma}\left(\omega,t_{\mathrm{pr}}\right)=\sum_{\boldsymbol{k}}\boldsymbol{\sigma}_{\boldsymbol{k}}\left(\omega,t_{\mathrm{pr}}\right)$,
each single $\boldsymbol{\sigma}_{\boldsymbol{k}}\left(\omega,t_{\mathrm{pr}}\right)$
can have odd-pump-frequency components that come from the coupling
via the odd-terms in the expansion of the Peierls substitution which
are proportional to the velocity and higher order odd derivatives
(see Ref.~\citep{eskandari2023dynamical}). Summing over all of the
\textbf{k} points on the grid these terms add up to zero because of
the periodicity of the FBZ. This cancels out all odd-pump-frequency
components in $\boldsymbol{\sigma}\left(\omega,t_{\mathrm{pr}}\right)$,
and consequently in the differential reflectivity.

\begin{figure*}[t]
\centering{}\includegraphics[width=12cm]{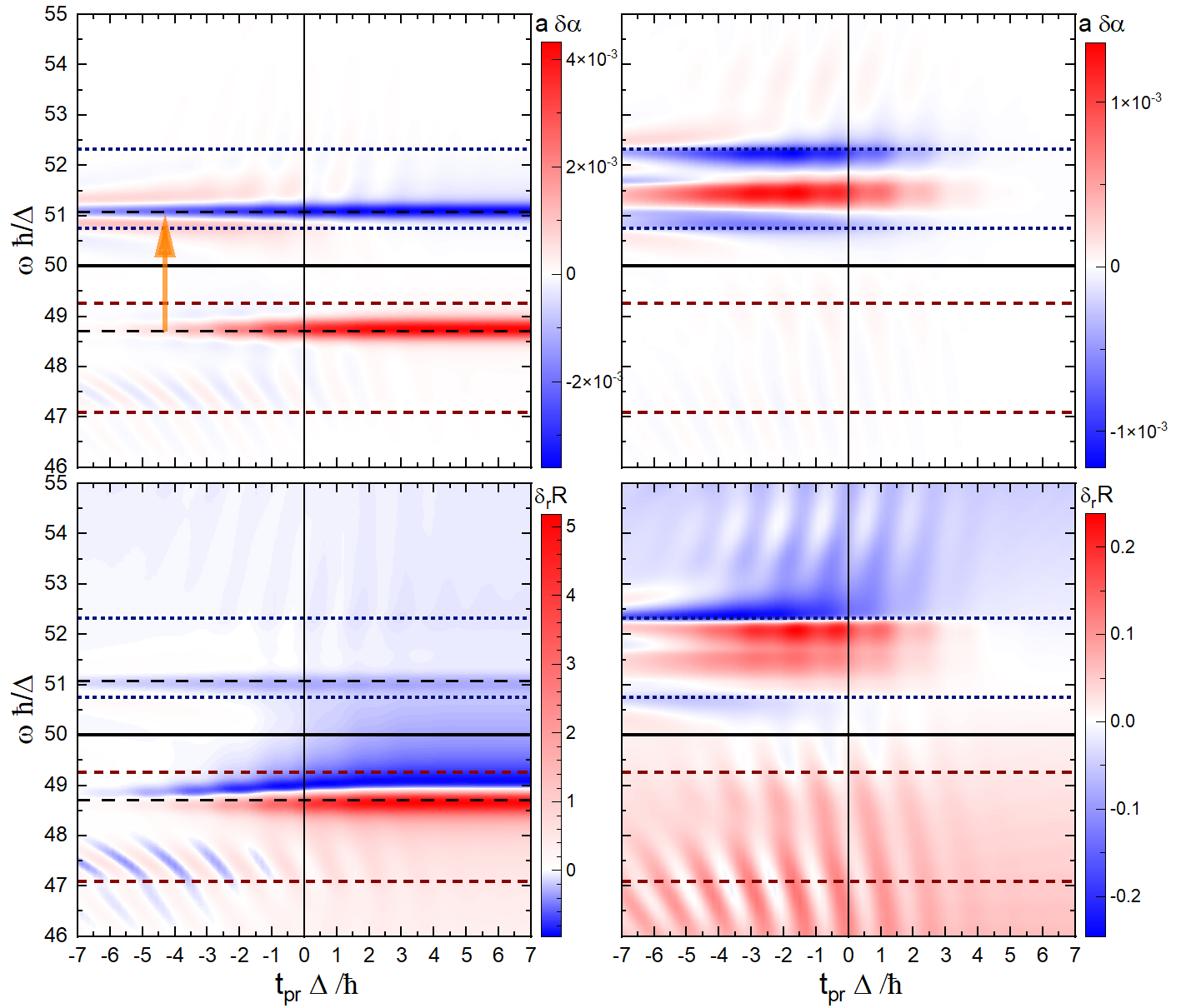}\caption{Transient (top row) differential absorption and (bottom row) relative
differential reflectivity, with (left column) only inter-band and
(right column) only intra-band transitions in the dynamics, vs delay
time and probe photon energy. The horizontal lines are the same as
those of Fig.~\ref{fig:dR55_D=00003D0}. \protect\label{fig:dR55_D=00003D0_inter_intra}}
\end{figure*}

\subsection{Inter and intra-band transitions\protect\label{subsec:Inter-and-intra-band}}

In Fig.~\ref{fig:dR55_D=00003D0_inter_intra}, we study the separate
effects of the so-called inter- and intra-band transitions in the
dynamics. To do that, one computes the projection coefficients, $P_{\boldsymbol{k}nn^{\prime}}\left(t\right)$,
from the Eq.~\ref{eq:P-dyn-gen}, by considering just either the
inter- or the intra-band transitions in the dynamics (see Ref.~\citep{eskandari2023dynamical}
for a detailed explanation). Fig.~\ref{fig:dR55_D=00003D0_inter_intra}
(top-left panel) shows the transient differential absorption coefficient
with only the inter-band transitions active. In this case, the energies
where we have one-photon resonances (the one-photon-resonance energies)
are even more relevant than in the case of full dynamics (Fig.~\ref{fig:dR55_D=00003D0}(a)),
as the off-resonance energies become less relevant without the intra-band
transitions.

Fig.~\ref{fig:dR55_D=00003D0_inter_intra} (top-right panel) shows
the transient differential absorption coefficient with only the intra-band
transitions active. In this case, the resonant energy gaps are no
longer relevant because there is no electron transition among the
bands, so the resonance loses meaning. Given that with our system
and pump parameters, the resonance effects are the most relevant ones
(see Fig.~\ref{fig:dR55_D=00003D0}(a)), this may lead to the conclusion
that the intra-band motion is irrelevant. The qualitative shape of
the color map, i.e., the photon energies at which positive or negative
(red or blue, respectively) signals appear, is almost independent
of the pump frequency (see Fig.~\ref{fig:dR55_D=00003D0_wps-intra}
in App.~\ref{sec:Intra-band}), even though the details of the oscillations
depend on it. Consequently, the main features for this case are determined
by the electronic bands and their couplings to the pump pulse (i.e.,
to the system properties) rather than to the pump-pulse parameters.
A comparison between the full-dynamics case and the cases for only
inter- and intra-band transitions active reveals that each type of
transition contributes with specific features, but that only the interplay
between them can explain the full-dynamics case. The features caused
by inter-band transitions are the dominant ones. The lower relevance
of intra-band transitions is due to the smoothness of the bands caused
by the short range and the relatively small values of the hoppings.
This leads to relatively small changes in the bands upon shifting
the momentum by the vector potential through the Peierls substitution.

In Fig.~\ref{fig:dR55_D=00003D0_inter_intra} (bottom-left/bottom-right
panel), we have the transient relative differential reflectivity for
the case of having only inter-/intra-band transitions in the dynamics.
Similar to the case of absorption, we see that there are features
in the full-dynamics case originating from either of the inter- or
intra-band transitions and the interplay of the two types of transitions
results in the full picture (compare with Fig.~\ref{fig:dR55_D=00003D0}(c)).

\begin{figure*}[t]
\centering{}\includegraphics[width=17cm]{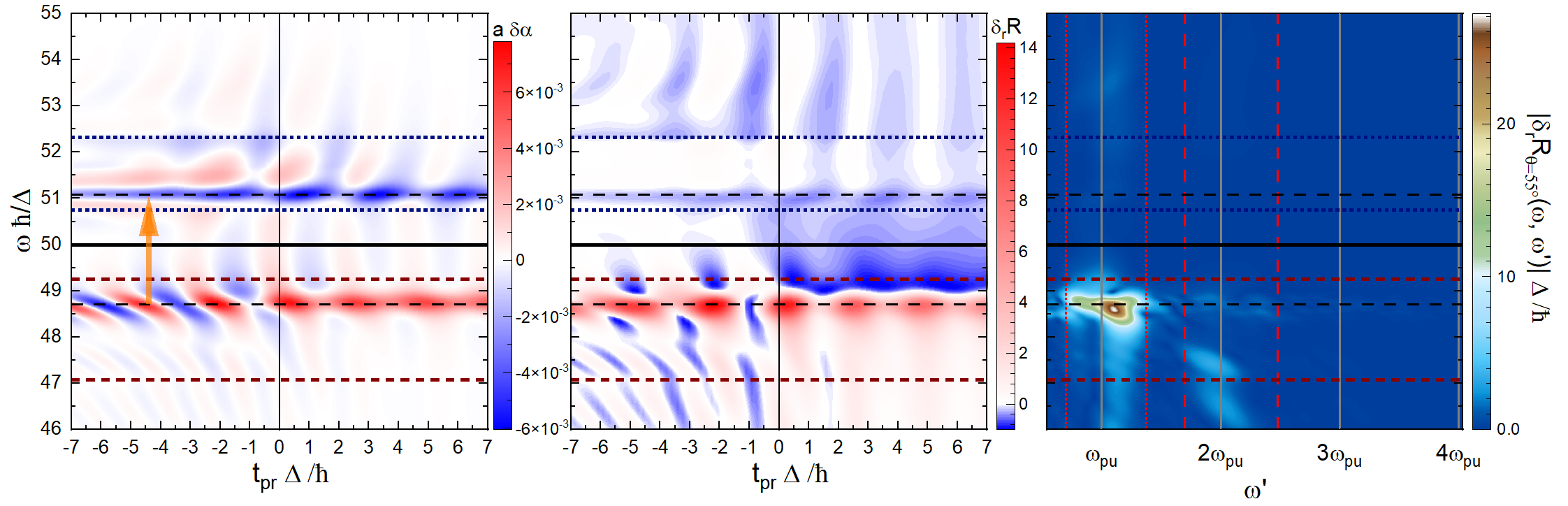}\caption{Transient optical behavior of the pumped system with both the local
dipole and the Peierls substitution coupling with the pump. (left)
The transient differential absorption coefficient, as given by Eq.~\ref{eq:dalpha},
and, (middle) transient relative differential reflectivity at the
incident angle of $\theta=55{^\circ}$, vs delay time and probe photon
energy. (right) The amplitude of the Fourier transformation of the
transient reflectivity, $\left|\delta R_{\theta=55{^\circ}}\left(\omega,\omega^{\prime}\right)\right|$,
see Eq.~\ref{eq:FTdR} (we don't show the low frequency region, $\omega^{\prime}<0.2\Delta/\hbar$)
. The horizontal lines are the same as those of Fig.~\ref{fig:dR55_D=00003D0}.
\protect\label{fig:dR55_D=00003D0.05}}
\end{figure*}

\subsection{Dipole coupling\protect\label{subsec:Dipole-coupling}}

Now, we study the effect of having a finite dipole, in addition to
the Peierls substitution coupling. Although the theory has been developed
and reported in the previous sections without any restriction on the
momentum dependence of the dipole, to keep the toy model simple, we
consider local dipole elements $\tilde{\mathbf{D}}_{\mathbf{R}=\boldsymbol{0},1,2}=\tilde{\mathbf{D}}_{\mathbf{R}=\boldsymbol{0},2,1}^{*}=\mathrm{i}0.05a\hat{\mathbf{j}}$,
which obviously results in a momentum-independent dipole.

In Fig.~\ref{fig:dR55_D=00003D0.05} (top-left panel), we show the
transient differential absorption coefficient for such a case. Similar
to the former case, the main change in the absorption occurs at the
one-photon resonance energies in the VB and CB. In this case, the
changes in the absorption are higher as the local dipole strengthens
the excitation rates of electrons from the VB to CB.

In Fig.~\ref{fig:dR55_D=00003D0.05} (middle panel), we show the
transient relative differential reflectivity at the incident angle
$\theta=55{^\circ}$, $\delta R_{\theta=55{^\circ}}\left(\omega,t_{\mathrm{pr}}\right)$.
We see that the \emph{white lines} that determine the absorption edges
are somewhat distorted under the effect of the local dipole coupling.
This is due to the off-resonant non-permanent-excitation oscillations
at other energies that are strengthened by the local dipole and can
have more long-range tails in energy.

Comparison between Figs.~\ref{fig:dR55_D=00003D0} and \ref{fig:dR55_D=00003D0.05}
clarifies that the oscillations are different in the two cases. This
can be well understood from Fig.~\ref{fig:dR55_D=00003D0.05} (right
panel), where we show the amplitude of the Fourier transformation
of the transient reflectivity, $\left|\delta R_{\theta=55{^\circ}}\left(\omega,\omega^{\prime}\right)\right|$.
The main difference with the former case is that here we do have the
odd-pump-frequency components, and in particular, $\omega^{\prime}$
around 1$\omega_{\mathrm{pu}}$ giving the most substantial contribution
in the oscillatory behavior. The odd-pump-frequency components come
from the coupling to the pump-pulse electric field via the local dipole.
Such contributions are not canceled upon the summation over the entire
k grid, unlike what happens for the velocity and higher-order odd-derivatives
in the Peierls substitution, and instead significantly contribute
to the system's response.

In the following calculations, we consider zero dipole coupling between
the VB and CB. Another point that is worthy of being mentioned is
that the 2$\omega_{\mathrm{pu}}$-component is different from the
case of having no local dipole, Fig.~\ref{fig:dR55_D=00003D0}. It
shows the interplay between the two terms, which clarifies that one
cannot consider their effects irrespective of each other.

\begin{figure*}[t]
\centering{}\includegraphics[width=18cm]{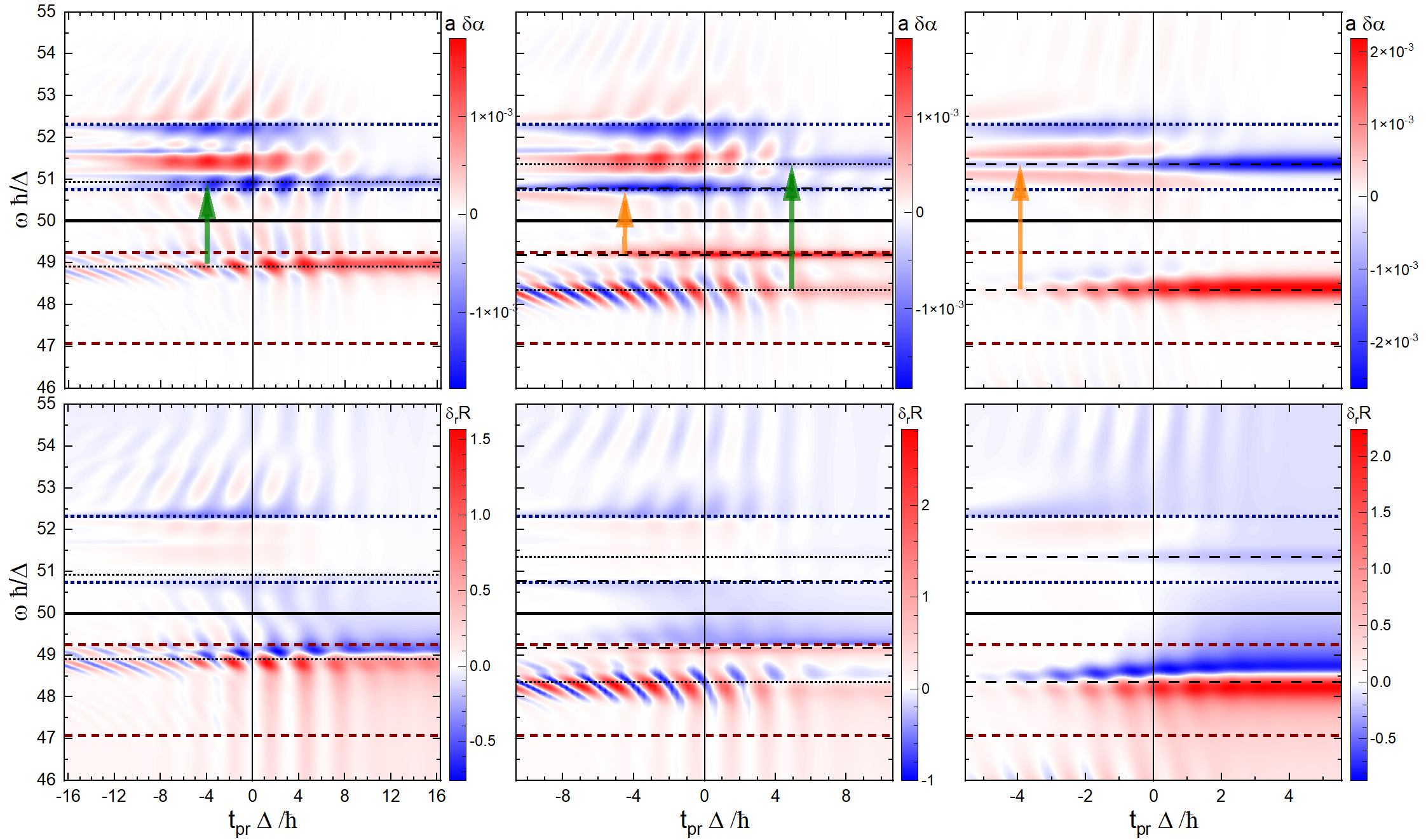}\caption{(top row) Transient differential absorption coefficients and (bottom
row) transient relative differential reflectivities, vs delay time
and probe photon energy, for three different pump-pulse frequencies:
(left column) $\hbar\omega_{\mathrm{pu}}/\Delta=1$, (middle column)$\hbar\omega_{\mathrm{pu}}/\Delta=1.5$
and (right column) $\hbar\omega_{\mathrm{pu}}/\Delta=3$. For each
$\omega_{\mathrm{pu}}$, the FWHM of the pump pulse,$\tau_{\mathrm{pu}}$,
is set in such a way that $\omega_{\mathrm{pu}}\tau_{\mathrm{pu}}$
remains the same as the one of Fig.~\ref{fig:dR55_D=00003D0}. The
orange (green) arrows indicate the one-photon (two-photon) transitions,
if any, while the horizontal black-dashed (-dotted) lines determine
the peaks of one-photon (two-photon) DORS, if any. \protect\label{fig:dR55_D=00003D0_wps}}
\end{figure*}

\subsection{Pump-pulse frequency dependence\protect\label{subsec:Pump-pulse-frequency-dependence}}

In Fig.~\ref{fig:dR55_D=00003D0_wps}, we study the effect of changing
the pump-pulse frequency to better understand the role of resonances
with the pump-pulse frequency in the out-of-equilibrium optical properties.
The FWHM of the pump pulse, $\tau_{\mathrm{pu}}$, is also changed
in such a way that, in all three cases, $\omega_{\mathrm{pu}}\tau_{\mathrm{pu}}$
remains the same as the one of Fig.~\ref{fig:dR55_D=00003D0}. In
Fig.~\ref{fig:dR55_D=00003D0_wps} (top-left panel), we show the
transient differential absorption coefficient for the case of $\hbar\omega_{\mathrm{pu}}/\Delta=1$.
It is worth noting that the minimum gap energy in our system, which
is at $\boldsymbol{\Gamma}$, is $1.5\Delta$, which is higher than
$\hbar\omega_{\mathrm{pu}}$, and therefore we have no one-photon
resonances. However, this map indicates a main absorption process
at the two-photon resonant \textbf{k}-points (a green arrow indicates
the two-photon transitions) at the energies where we have the peaks
of the two-photon DORS, as indicated by the black-dotted lines. The
process is of the second order; hence, the absorption change is lower
than in the former cases. The bottom-left panel shows the transient
relative differential reflectivity for the same pump-pulse parameters.
Again, the peaks of the two-photon DORS (signaled by black-dotted
lines) indicate the absorption edges, though they are not as intense
as the first-order transitions.

In Fig.~\ref{fig:dR55_D=00003D0_wps} (top-middle and bottom-middle
panels), we show the transient differential absorption coefficient
and relative differential reflectivity, respectively, for $\hbar\omega_{\mathrm{pu}}/\Delta=1.5$.
The pump-pulse frequency resonates with the minimum energy gap at
$\boldsymbol{\Gamma}$. Therefore, the upper/lower edges of the VB/CB
determine the one-photon-resonance energies and almost coincide with
the peaks of the one-photon DORS and, consequently, the main absorption
stripes. In addition to the one-photon resonances, we also see the
effects of the two-photon resonances, which are of the second order
(the peaks of the two-photon DORS are signaled by black-dotted lines,
and and a green arrow indicates the two-photon transitions). Both
first-order and second-order transitions result in their corresponding
absorption edges in the transient relative differential reflectivity,
with the difference that the absorption edges of the two-photon resonances
are less intense than the ones of the one-photon resonances.

In Fig.~\ref{fig:dR55_D=00003D0_wps} (top-right and bottom-right
panels), we show the transient differential absorption coefficient
and relative differential reflectivity, respectively, for $\hbar\omega_{\mathrm{pu}}/\Delta=3$.
In this case, we have only the one-photon resonances, with their corresponding
effects in the absorption and reflectivity, similar to the ones of
Fig.~\ref{fig:dR55_D=00003D0}. It is noteworthy that in this case,
because of the high value of $\omega_{\mathrm{pu}}$, we don't have
two-photon resonances at all. On the other hand, in the case of Fig.~\ref{fig:dR55_D=00003D0},
even though we do have two-photon resonances, they have much smaller
effects than the one-photon ones because they occur near the edges
of the VB and CB and, hence, (i) they are coupled very weakly to the
pump pulse (their velocity is small) and (ii) they have a smaller
DORS than the one-photon resonances (see also Ref. \citep{eskandari2023dynamical}).

\begin{figure*}[t]
\centering{}\includegraphics[width=12cm]{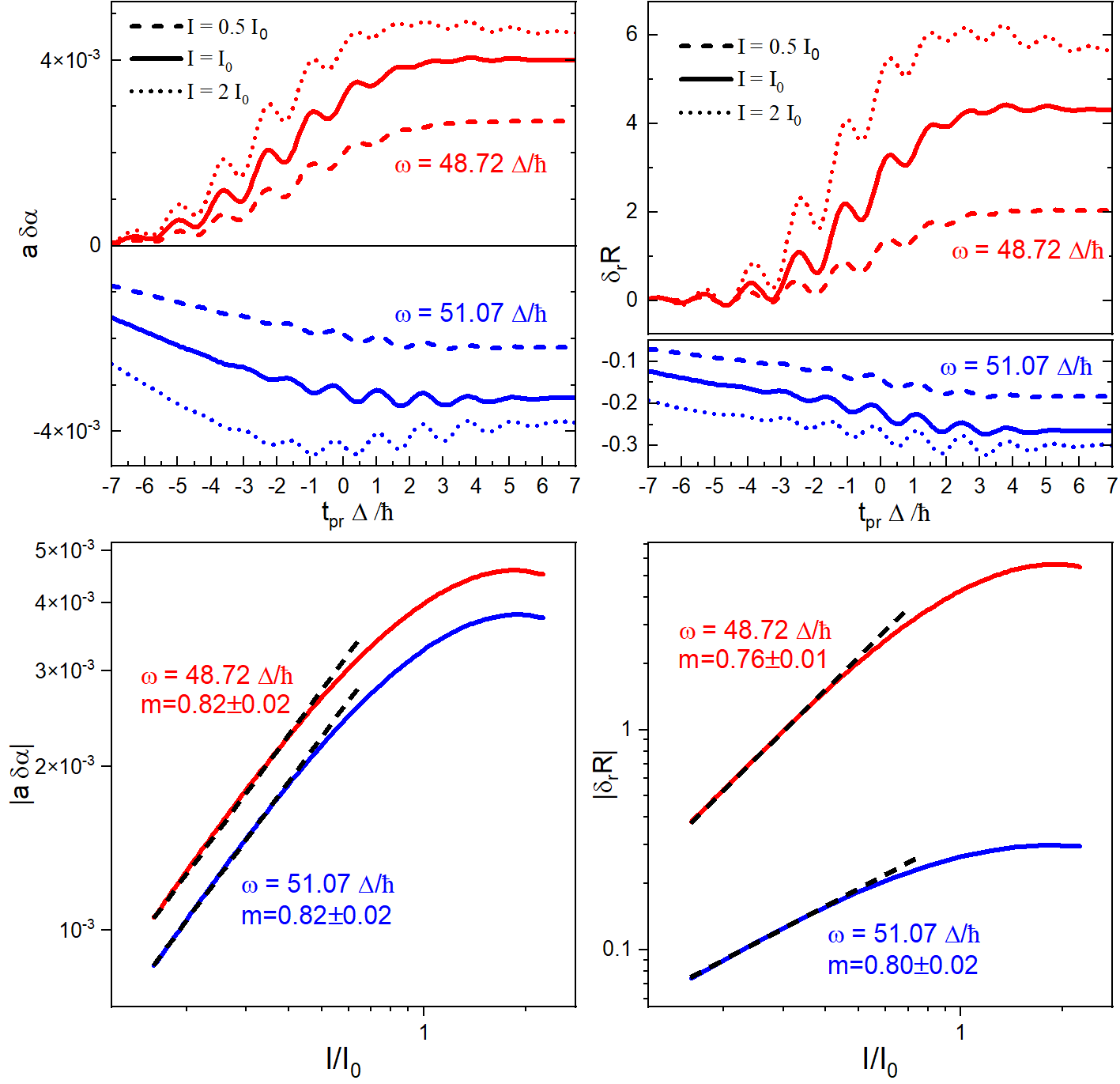}\caption{Study of the intensity dependence, for two probe frequencies: $\hbar\omega/\Delta=48.72$
and $51.07$. (top-left) The transient differential absorption coefficient,
and (top-right) the transient relative differential reflectivities,
vs the probe time, for three different pump-pulse intensities: $0.5I_{0}$,
$I_{0}$ and $2I_{0}$, where $I_{0}$ corresponds to the pump-pulse
intensity of Fig.~\ref{fig:dR55_D=00003D0} (other parameters are
also the same as the ones of Fig.~\ref{fig:dR55_D=00003D0}). The
absolute values of, (bottom-left) the residual differential absorption
coefficient, and (bottom-right) the residual relative differential
reflectivity, vs the pump-pulse intensity. The slope of each curve,
$m$, is written next to it. \protect\label{fig:Intensities}}
\end{figure*}

\subsection{Pump-pulse intensity dependence\protect\label{subsec:Pump-pulse-intensity-dependence}}

Another relevant study investigates the effect of the pump-pulse intensity.
The complete maps with different intensities look qualitatively similar
(not shown), and for a better understanding, we look at some specific
probe frequencies. For this purpose, we considered two probe frequencies
in Fig.~\ref{fig:Intensities}: $\hbar\omega/\Delta=48.72$ and $51.07$.
In the top panels of Fig.~\ref{fig:Intensities}, we consider three
different pump-pulse intensities: $0.5I_{0}$, $I_{0}$ and $2I_{0}$,
where $I_{0}$ corresponds to the pump-pulse intensity of Fig.~\ref{fig:dR55_D=00003D0}
(other parameters are also the same as the ones of Fig.~\ref{fig:dR55_D=00003D0}),
and plot the transient differential absorption coefficients, and the
transient relative differential reflectivities, vs. the probe time.
At each instant of time, one cannot find a power-law behavior for
either of the quantities. Nevertheless, the order is respected; generally
speaking, increasing the intensity increases the strength of the signals.

In the bottom panels of Fig.~\ref{fig:Intensities}, we plot the
absolute value of the residual differential absorption coefficient,
$\left|\delta\alpha\left(\omega,t_{\mathrm{pr}}\rightarrow+\infty\right)\right|$,
and the absolute value of the residual relative differential reflectivity,
$\left|\delta R_{\theta=55{^\circ}}\left(\omega,t_{\mathrm{pr}}\rightarrow+\infty\right)\right|$,
vs. the pump-pulse intensity (note that $t_{\mathrm{pr}}\rightarrow+\infty$
means a time well after the application of the pump pulse, but still
much smaller than the time scale of the other decoherence and de-excitation
transitions, such as electron-phonon interaction, spontaneous emission,
etc.). We see a linear behavior in the logarithmic plot for lower
intensities, which means a power-law dependence on the intensity.
The slope of the logarithmic plot, $m$, is the power of the dependence
and, in our cases, is always below 1. The one-photon resonance is
the main process in our model for $\omega_{\mathrm{pu}}=2.33\Delta/\hbar$.
But the power is less than 1, as all orders of the Peierls expansion
co-exist and affect each other (see Refs.~\citep{inzani2022field}
and \citep{eskandari2023dynamical} for more discussion). For higher
intensities, the curves bend and show no more a power-law behavior.
This can be explained by noting that the excitations are far below
the population inversion for low intensities, and by varying the intensity,
one can find a power law. Increasing the intensity increases the Rabi
frequency, which results in excitations closer to the full population
inversion. Hence, we do not find a power law behavior with respect
to the intensity anymore.

\begin{figure*}[t]
\centering{}\includegraphics[width=18cm]{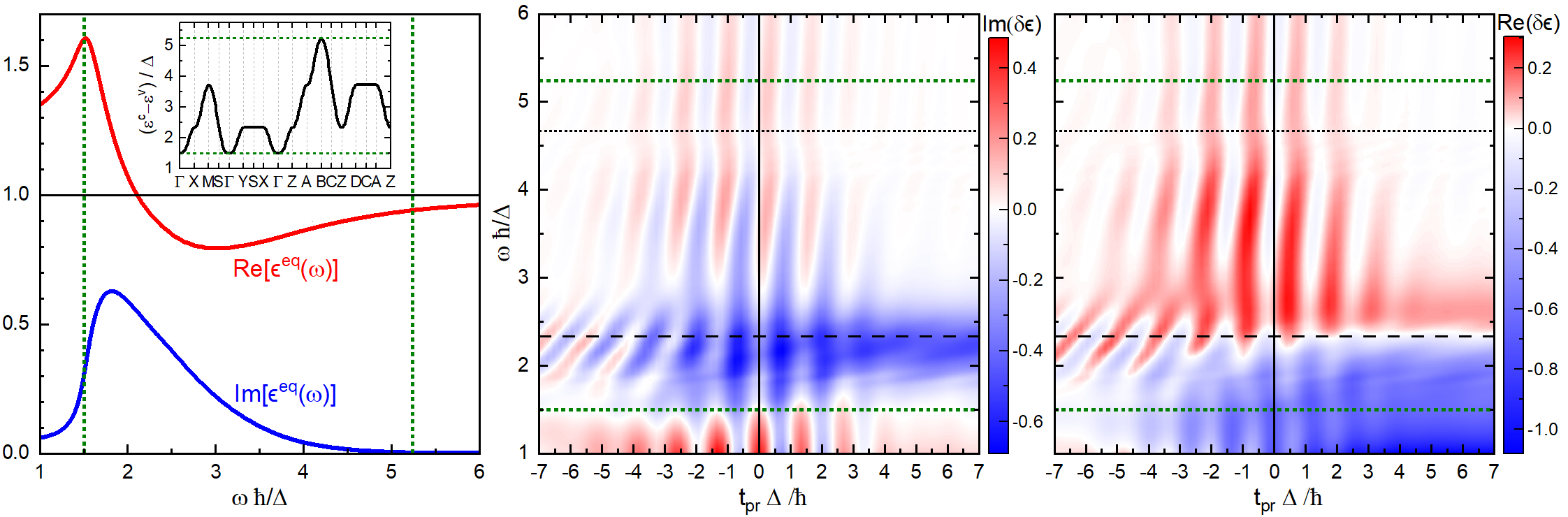}\caption{Optical properties in the IR-V regime. (left) The imaginary and real
parts of the dielectric function at equilibrium. The inset shows the
gap, $\varepsilon_{\boldsymbol{k}}^{c}-\varepsilon_{\boldsymbol{k}}^{v}$,
along the main path in the \textbf{k} space. The transient differential
(middle) imaginary and (right) real parts of the dielectric function.
The green-dotted lines determine the minimum and maximum values of
the gap energies and the black-dashed (-dotted) line show the energy
of one-(two-)photon resonance. \protect\label{fig:IRV}}
\end{figure*}

\subsection{IR and visible regime\protect\label{subsec:IR-and-visible}}

Up to now, we have been studying the probes in the high-frequency
regime of XUV. Now, we move to probes in the low-frequency regime,
which corresponds to IR and visible regimes, which we denote by IR-V.
Based on our arguments in App.~\ref{sec:out-of-eq-optics}, using
the standard formulas for the transient optical properties (reflectivity
and absorption) in this regime is not straightforward. Indeed, to
study the transient optical behavior, we can safely investigate the
transient differential imaginary and real parts of the dielectric
function (see also App.~\ref{sec:differential-epsilon}).

In Fig.~\ref{fig:IRV} (left panel), we show the imaginary and real
parts of the dielectric function at equilibrium in the IR-V regime.
In this regime, we use $0^{+}=0.1\Delta/\hbar$ (see App.~\ref{sec:zp_study},
Fig.~\ref{fig:zp_stud}). The \emph{core} bands play no role, and
the probe pulse induces transitions between the VB and CB. The green-dotted
lines determine the minimum and maximum values of the gap energies
and the CB-VB gap, $\varepsilon_{\boldsymbol{k}}^{c}-\varepsilon_{\boldsymbol{k}}^{v}$,
along the main path in the \textbf{k} space is shown in the inset.
$\Re\left[\epsilon^{\mathrm{eq}}\left(\omega\right)\right]$ decreases
outside of the energy range of the gap energies and tends to its predicted
value for $\omega\rightarrow0$ (see Eq.~\ref{eq:eps_w0}). The $\Im\left[\epsilon^{\mathrm{eq}}\left(\omega\right)\right]$
is finite only in the range of the gap energies, but because of the
finite value of $0^{+}$, it doesn't tend to zero for $\omega\rightarrow0$.
Instead, as shown in Eq.~\ref{eq:eps_w0}, it artificially diverges
in this limit (not shown).

Fig.~\ref{fig:IRV} (middle panel) shows the imaginary part of the
transient differential dielectric function. At the pump-pulse frequency,
we have the most relevant resonant excitation process. The electrons
get excited from the VB to CB, reducing the absorption of the probe-pulse
via the same process, as there remain fewer available electrons in
the VB and vacancies in the CB, hence, a bright blue stripe. In the
right panel of Fig.~\ref{fig:IRV}, we report the real part of the
transient differential dielectric function, which shows a clear absorption
edge at the resonance energy. It is noteworthy that, as we mentioned
before, the two-photon resonance is very weak in this case; hence,
there is almost no signature at its corresponding energy.

\section{Summary\protect\label{sec:Summary-and-conclusions}}

We provided the linear response theory to a weak probe pulse for a
pumped system out of equilibrium. Such a theory speeds up numerical
calculations enormously, as one needs to simulate the pumped system
only once instead of considering the probe repeatedly for all pump-probe
delays.

Applying this theory to a generic semiconductor system with a quadratic
Hamiltonian in the dipole gauge and using the recently developed theory,
Dynamical Projective Operatorial Approach, we derive a formula for
computing the optical conductivity as a function of the pump-probe
delay time and the probe photon frequency, from which one can obtain
the transient reflectivity and absorption. We provided several essential
guidelines that would make it possible to perform actual numerical
calculations with affordable computational costs.

By considering a prototypical three-band (valence, conduction, and
core) model system, which is pumped in the frequency range that can
excite electrons between valence and conduction bands (which is assumed
to be the IR range), we computed transient optical properties in the
probe photon frequency range of the gaps between the core band and
the other bands (which is assumed to be the XUV range). We provided
a systematic way to analyze such results to find the relation between
the features in the transient differential absorption and reflectivity
and the system band structure in the first Brillouin zone. We also
proposed and exploited some generalizations of the density of states
related solely to the resonant states. Moreover, we studied the effect
of a momentum-independent dipole and different mechanisms, such as
inter-band and intra-band transitions and single- and multi-photon
processes. The latter is further analyzed by changing the photon frequency
to explore various regions of the first Brillouin zone in terms of
being in resonance with the pump pulse. Furthermore, we studied the
effect of the pump pulse intensity. Finally, we investigated the transient
optical response for the IR and visible range and analyzed the results
and their relations to the system and pump pulse features.

Our work is at least of two-fold relevance: (i) we provide a theoretical
method that can be affordably applied to simulate realistic experimental
setups, and (ii) we give a systematic way to analyze the results of
optical measurements in terms of the system properties and pump pulse
parameters.
\begin{acknowledgments}
The authors thank Claudio Giannetti, Matteo Lucchini, Stefano Pittalis,
and Carlo Andrea Rozzi for the insightful discussions. The authors
acknowledge support by MIUR under Project No. PRIN 2017RKWTMY.
\end{acknowledgments}

\appendix

\section{The linear variation of the density matrix in the presence of the
probe pulse\protect\label{sec:DM-linear-variation}}

The time evolution of the density matrix of the pumped system, $\hat{\Upsilon}^{\prime}\left(t\right)$,
is given by the following equation of motion
\begin{equation}
\mathrm{i}\hbar\frac{\partial}{\partial t}\hat{\Upsilon}^{\prime}\left(t\right)=\left[\hat{H}^{\prime}\left(t\right),\hat{\Upsilon}^{\prime}\left(t\right)\right],
\end{equation}
with generalized solution

\begin{equation}
\hat{\Upsilon}^{\prime}\left(t\right)=\hat{U}^{\prime}\left(t,t_{\mathrm{ini}}\right)\hat{\Upsilon}_{0}\hat{U}^{\prime}\left(t_{\mathrm{ini}},t\right),\label{eq:DM_p}
\end{equation}
where $\hat{\Upsilon}_{0}$ is the density matrix at the time $t_{\mathrm{ini}}\rightarrow-\infty$
(i.e., at a time preceding the application of the pump) and,
\begin{equation}
\hat{U}^{\prime}\left(t_{1},t_{2}\right)=T_{+}\left[\mathrm{e}^{-\frac{\mathrm{i}}{\hbar}\int_{t_{2}}^{t_{1}}dt^{\prime}\hat{H}^{\prime}\left(t^{\prime}\right)}\right],\label{eq:U_p}
\end{equation}
 is the time propagator in which $T_{+}$ is the time-ordering operator.

The full density matrix of the system in the presence of both the
pump and probe pulses, $\hat{\Upsilon}\left(t\right)$, obeys the
following equation of motion,

\begin{equation}
\mathrm{i}\hbar\frac{\partial}{\partial t}\hat{\Upsilon}\left(t\right)=\left[H\left(t\right),\hat{\Upsilon}\left(t\right)\right].
\end{equation}
We need to consider the full effect of the pumping via $\hat{H}^{\prime}\left(t\right)$,
but consider the effect of the probe pulse up to the linear order
in $\hat{H}^{\prime\prime}\left(t\right)$. In order to do that, we
move to an interaction picture in which the Hamiltonian of the probe
pulse, $\hat{H}^{\prime\prime}\left(t\right)$, is considered as the
interaction term. The operators in this picture are indicated by the
subscript $\hat{H}^{\prime}$ and given by Eq.~\ref{eq:Hp_S_trans}.
The full density matrix in this picture is given by,

\begin{equation}
\hat{\varUpsilon}_{H^{\prime}}\left(t\right)=\hat{U}^{\prime}\left(t_{\mathrm{ini}},t\right)\hat{\Upsilon}\left(t\right)\hat{U}^{\prime}\left(t,t_{\mathrm{ini}}\right).\label{eq:DM_Hp_1}
\end{equation}
It is straightforward to perform a direct time derivation and show
that $\hat{\varUpsilon}_{H^{\prime}}\left(t\right)$ obeys the following
equation of motion,

\begin{align}
\mathrm{i}\hbar\frac{\partial}{\partial t}\hat{\varUpsilon}_{H^{\prime}}\left(t\right) & =\left[\hat{H}_{H^{\prime}}^{\prime\prime}\left(t\right),\hat{\varUpsilon}_{H^{\prime}}\left(t\right)\right],
\end{align}
with generalized solution
\begin{equation}
\hat{\varUpsilon}_{H^{\prime}}\left(t\right)=\hat{U}^{\prime\prime}\left(t,t_{\mathrm{ini}}\right)\hat{\Upsilon}_{0}\hat{U}^{\prime\prime}\left(t_{\mathrm{ini}},t\right),\label{eq:DM_Hp_2}
\end{equation}
where
\begin{equation}
\hat{U}^{\prime\prime}\left(t_{1},t_{2}\right)=T_{+}\left[e^{-\frac{i}{\hbar}\int_{t_{2}}^{t_{1}}dt^{\prime}\hat{H}_{H^{\prime}}^{\prime\prime}\left(t^{\prime}\right)}\right].
\end{equation}

Combining Eqs.~\ref{eq:DM_p}, \ref{eq:DM_Hp_1} and \ref{eq:DM_Hp_2},
the probe-pulse-induced variation of the density matrix is obtained
as,
\begin{equation}
\hat{\Upsilon}\left(t\right)-\hat{\Upsilon}^{\prime}\left(t\right)=\hat{U}^{\prime}\left(t,t_{\mathrm{ini}}\right)\left(\hat{\varUpsilon}_{H^{\prime}}\left(t\right)-\hat{\Upsilon}_{0}\right)\hat{U}^{\prime}\left(t_{\mathrm{ini}},t\right).\label{eq:var_DM_1}
\end{equation}
Up to the linear order in $\hat{H}^{\prime\prime}\left(t\right)$,
we have
\begin{equation}
\hat{U}^{\prime\prime}\left(t_{1},t_{2}\right)\simeq1-\frac{i}{\hbar}\int_{t_{2}}^{t_{1}}dt^{\prime}\hat{H}_{H^{\prime}}^{\prime\prime}\left(t^{\prime}\right).\label{eq:Upp_Lin}
\end{equation}
Substituting Eq.~\ref{eq:Upp_Lin} in Eq.~\ref{eq:DM_Hp_2} and
then in Eq.~\ref{eq:var_DM_1}, we obtain the linear order variation
of the density matrix, as given in Eq.~\ref{eq:DM_LR} in the main
text.

\section{Dielectric function and the reflectivity formula out of equilibrium\protect\label{sec:out-of-eq-optics}}

\subsection{Dielectric function}

In order to obtain the dielectric function from the conductivity out
of equilibrium, we recall that the continuity equation for $\boldsymbol{j}_{\mathrm{pr}}$,
the electrical current induced by the probe pulse, reads as

\begin{equation}
\nabla\cdot\boldsymbol{j}_{\mathrm{pr}}\left(t\right)+\partial_{t}\rho_{\mathrm{pr}}\left(t\right)=0,\label{eq:cont_pr}
\end{equation}
where $\rho_{\mathrm{pr}}\left(t\right)$ is the bounded charge density
induced only by the probe pulse, and for the sake of simplicity, we
showed only the time dependence explicitly. The probe pulse induced
quantities (such as current, charge density, etc.) are obtained as
the difference between the value of the quantities in the presence
of both pump and probe pulses and their value in the presence of the
pump pulse only.

The bounded charge density satisfies the Gauss's law, as
\begin{equation}
\nabla\cdot\left[\epsilon_{0}\boldsymbol{E}_{\mathrm{pr}}\left(t\right)-\boldsymbol{\mathcal{D}}_{\mathrm{pr}}\left(t\right)\right]=\rho_{\mathrm{pr}}\left(t\right),\label{gauss_pr}
\end{equation}
where $\boldsymbol{\mathcal{D}}_{\mathrm{pr}}\left(t\right)$ is the
electric displacement field of the probe pulse, i.e., it is constructed
from the electric field of the probe pulse and the probe-pulse induced
charge polarization. Combining Eqs.~\ref{eq:cont_pr} and \ref{gauss_pr}
and considering the boundary conditions, we arrive at the following
formula,
\begin{equation}
\boldsymbol{j}_{\mathrm{pr}}\left(t\right)+\epsilon_{0}\partial_{t}\boldsymbol{E}_{\mathrm{pr}}\left(t\right)-\partial_{t}\boldsymbol{\mathcal{D}}_{\mathrm{pr}}\left(t\right)=0.\label{eq:j_E_D}
\end{equation}
The responses to the probe pulse are assumed to be linear, so that
one can write
\begin{equation}
\boldsymbol{j}_{\mathrm{pr}}\left(t\right)=\int dt^{\prime}\boldsymbol{\sigma}\left(t,t^{\prime}\right)\cdot\boldsymbol{E}_{\mathrm{pr}}\left(t^{\prime}\right),\label{eq:J_sig_E_gen}
\end{equation}
and
\begin{equation}
\boldsymbol{\mathcal{D}}_{\mathrm{pr}}\left(t\right)=\epsilon_{0}\int dt^{\prime}\boldsymbol{\epsilon}\left(t,t^{\prime}\right)\cdot\boldsymbol{E}_{\mathrm{pr}}\left(t^{\prime}\right),\label{eq:D_eps_E}
\end{equation}
where $\boldsymbol{\epsilon}\left(t,t^{\prime}\right)$ is the (dimensionless)
dielectric function of the material and $\epsilon_{0}$ is the vacuum
permeability. Substituting back in Eq.~\ref{eq:j_E_D}, performing
a Fourier transformation with respect to the time $\left(t-t^{\prime}\right)$,
and using integration by parts to deal with the time derivatives,
we get
\begin{multline}
\int dt^{\prime}\boldsymbol{\epsilon}\left(\omega,t^{\prime}\right)\cdot e^{i\omega t^{\prime}}\boldsymbol{E}_{\mathrm{pr}}\left(t^{\prime}\right)=\\
\int dt^{\prime}e^{i\omega t^{\prime}}\boldsymbol{E}_{\mathrm{pr}}\left(t^{\prime}\right)+\frac{i}{\omega\epsilon_{0}}\int dt^{\prime}\boldsymbol{\sigma}\left(\omega,t^{\prime}\right)\cdot e^{i\omega t^{\prime}}\boldsymbol{E}_{\mathrm{pr}}\left(t^{\prime}\right).
\end{multline}
Note that the response functions $\boldsymbol{\sigma}\left(\omega,t^{\prime}\right)$
and $\boldsymbol{\epsilon}\left(\omega,t^{\prime}\right)$ depend
only on the material and the pump pulse, and the above equation should
hold for any probe pulse, $\boldsymbol{E}_{\mathrm{pr}}\left(t^{\prime}\right)$.
Consequently, one can eliminate $\int dt^{\prime}e^{i\omega t^{\prime}}\boldsymbol{E}_{\mathrm{pr}}\left(t^{\prime}\right)$
and arrive at the following relation which computes the dielectric
function from the conductivity: 
\begin{equation}
\boldsymbol{\epsilon}\left(\omega,t\right)=\boldsymbol{1}+\frac{i}{\omega\epsilon_{0}}\boldsymbol{\sigma}\left(\omega,t\right).
\end{equation}

\subsection{General probe pulse}

In the derivation given in the main text, following the standard Green's
function approach for differential equations, we considered the electric
field of the probe pulse to have a Dirac-delta shape. Here, we explicitly
show that this does not cause any loss of generality on the results
of the derivations of the optical conductivity and the dielectric
function. Let $\boldsymbol{E}_{\mathrm{pr}}\left(t^{\prime}\right)$
be the electric field of a general probe pulse. It can be expanded
in terms of Dirac-delta functions as follows,
\begin{equation}
\boldsymbol{E}_{\mathrm{pr}}\left(t^{\prime}\right)=\int dt^{\prime\prime}\boldsymbol{E}_{\mathrm{pr}}\left(t^{\prime\prime}\right)\delta\left(t^{\prime}-t^{\prime\prime}\right).
\end{equation}
Being the probe pulse inherently weak, the total induced electric
current is the linear superposition of the electric currents induced
by each $\boldsymbol{E}_{\mathrm{pr}}\left(t^{\prime\prime}\right)\delta\left(t^{\prime}-t^{\prime\prime}\right)$
component. As a function of time $t$, the related induced current
is just $\boldsymbol{\sigma}\left(t,t^{\prime\prime}\right)\cdot\boldsymbol{E}_{\mathrm{pr}}\left(t^{\prime\prime}\right)$,
where $\boldsymbol{\sigma}\left(t,t^{\prime\prime}\right)$ is the
optical conductivity obtained in the main text (Eq.~\ref{eq:sigma-P-t}),
and hence the total current is given by
\begin{equation}
\boldsymbol{j}_{\mathrm{pr}}\left(t\right)=\int dt^{\prime\prime}\boldsymbol{\sigma}\left(t,t^{\prime\prime}\right)\cdot\boldsymbol{E}_{\mathrm{pr}}\left(t^{\prime\prime}\right).\label{eq:J_sigma_E_pp}
\end{equation}
Comparing Eqs.~\ref{eq:J_sig_E_gen} and \ref{eq:J_sigma_E_pp},
changing the dummy variable $t^{\prime\prime}\rightarrow t^{\prime}$,
and considering the arbitrariness of $\boldsymbol{E}_{\mathrm{pr}}\left(t^{\prime}\right)$,
one immediately concludes that the general optical conductivity is
the same as the one obtained using a Dirac-delta probe field, and
the latter choice does not cause any loss of generality to the final
result.

\subsection{Reflectivity}

Starting from the Maxwell's equations, after some straightforward
algebra and using Eq.~\ref{eq:D_eps_E} we obtain
\begin{align}
\nabla^{2}\boldsymbol{E}_{\mathrm{pr}}\left(\omega\right)= & -\mu_{0}\epsilon_{0}\omega^{2}\int dt^{\prime}e^{i\omega t^{\prime}}\boldsymbol{\epsilon}\left(\omega,t^{\prime}\right)\cdot\boldsymbol{E}_{\mathrm{pr}}\left(t^{\prime}\right),
\end{align}
and
\begin{align}
\nabla^{2}\boldsymbol{B}_{\mathrm{pr}}\left(\omega\right) & =-i\mu_{0}\epsilon_{0}\omega\int dt^{\prime}e^{i\omega t^{\prime}}\boldsymbol{\epsilon}\left(\omega,t^{\prime}\right)\cdot\partial_{t^{\prime}}\boldsymbol{B}_{\mathrm{pr}}\left(t^{\prime}\right),
\end{align}
where $\boldsymbol{B}_{\mathrm{pr}}$ is the magnetic field of the
probe pulse and we have considered a non-magnetic material, so that
the magnetic susceptibility of the material is just the one of vacuum,
$\mu_{0}$.

To proceed further, we need to apply some approximations. It is noticeable
that for the majority of the experimental setups, performing a Fourier
transformation of $\boldsymbol{\epsilon}\left(\omega,t^{\prime}\right)$
with respect to $t^{\prime}$, the highest frequency content would
be around $n\omega_{\mathrm{pu}}$ , with usually at most $n\approx2$.
If the width of the probe envelope is much less than the oscillation
period of the pump pulse, one can approximately take $\boldsymbol{\epsilon}\left(\omega,t^{\prime}\right)$
out of the above integrals. Given that the probe pulse oscillation
period should be smaller than the width of its envelope, our approximation
holds only when the probe pulse frequency is much higher than the
one of the pump pulse. Under such circumstances, one can write:
\begin{align}
\nabla^{2}\boldsymbol{E}_{\mathrm{pr}}\left(\omega\right)= & -\mu_{0}\epsilon_{0}\omega^{2}\boldsymbol{\epsilon}\left(\omega,t_{\mathrm{pr}}\right)\cdot\boldsymbol{E}_{\mathrm{pr}}\left(\omega\right),\label{eq:D2_eps_E}
\end{align}
and
\begin{align}
\nabla^{2}\boldsymbol{B}_{\mathrm{pr}}\left(\omega\right) & =-\mu_{0}\epsilon_{0}\omega^{2}\boldsymbol{\epsilon}\left(\omega,t_{\mathrm{pr}}\right)\cdot\boldsymbol{B}_{\mathrm{pr}}\left(\omega\right),\label{eq:D2_eps_B}
\end{align}
where $t_{\mathrm{pr}}$ is the center of the probe pulse and the
integration by parts has been used in the equation for the magnetic
field.

Eqs.~\ref{eq:D2_eps_E} and \ref{eq:D2_eps_B} ensure that, under
the approximations we have considered, the probe pulse behaves similarly
to the equilibrium case with the dielectric function given by $\boldsymbol{\epsilon}\left(\omega,t_{\mathrm{pr}}\right)$.
Consequently, one can apply the same procedure as at equilibrium and
obtain the formula for reflectivity given in Eq.~\ref{eq:Refl}.

\section{Considering only the transitions from the core levels in optical
conductivity\protect\label{sec:core-surface-cond}}

It is possible to assume that the high-frequency XUV probes detect
only the gaps between core bands and those bands near the Fermi energy
(either valence or conduction, which we call \textit{surface} bands,
and indicate by $n_{s}$). This approximation works better for larger
core-surface energy gaps. After some straightforward calculations,
one can show that the optical conductivity within this approximation
becomes

\begin{align}
 & \boldsymbol{\sigma}\left(t,t_{\mathrm{pr}}\right)\approx\frac{e}{\hbar\mathcal{V}}\theta\left(t-t_{\mathrm{pr}}\right)\nonumber \\
 & \sum_{\boldsymbol{k}}\int_{t_{\mathrm{pr}}}^{t}dt^{\prime}\sum_{n_{s}n_{s}^{\prime}n_{s}^{\prime\prime}n_{\mathrm{core}}}\mathrm{Im}\left[\boldsymbol{J}_{\boldsymbol{k}}^{n_{s}n_{\mathrm{core}}}\left(t\right)\right.\nonumber \\
 & \left.\left(-\frac{e}{\hbar}\boldsymbol{V}_{\boldsymbol{k}}^{n_{\mathrm{core}}n_{s}^{\prime}}\left(t^{\prime}\right)+e\boldsymbol{D}_{\boldsymbol{k}}^{n_{\mathrm{core}}n_{s}^{\prime}}\left(t_{\mathrm{pr}}\right)\delta\left(t^{\prime}-t_{\mathrm{pr}}\right)\right)\right.\nonumber \\
 & \left.e^{-i\omega_{\boldsymbol{k}n_{\mathrm{core}}}\left(t-t^{\prime}\right)}P_{\boldsymbol{k}n_{s}n_{s}^{\prime\prime}}^{\star}\left(t\right)P_{\boldsymbol{k}n_{s}^{\prime}n_{s}^{\prime\prime}}\left(t^{\prime}\right)\left(1-f_{\boldsymbol{k}n_{s}^{\prime\prime}}\right)\right].\label{eq:sigma-cores}
\end{align}
Unfortunately, the equilibrium conductivity computed via this approximation
has a non vanishing $1/\left(\omega+i0^{+}\right)$ tail. This can
be understood from Eq.~\ref{eq:sigma-eq-tail-2} as it has no core-surface
term, unlike Eq.~\ref{eq:sigma-eq-tail-1}. For the cases like our
toy-model this doesn't bring any problem, and the results of this
approximation are almost exactly the same as the ones obtained from
the full formula. However, in realistic situations this approximation
may fail, not only because of the non-vanishing tail, but also for
the effects of surface-surface transitions even on the conductivity
in XUV frequency range.

\section{Cancellation of the equilibrium tails at zero temperature\protect\label{sec:Cancellation-of-tails}}

We consider only the diagonal elements of the optical conductivity.
Accordingly, we use $\sigma=\boldsymbol{u}\cdot\boldsymbol{\sigma}\cdot\boldsymbol{u}$
where $\boldsymbol{u}$ is the unit vector of a generic direction.
The diagonal second tail term in equilibrium, Eq.~\ref{eq:sigma-eq-tail-2},
can be rewritten as

\begin{multline}
\sigma_{2}^{\mathrm{eq},\mathrm{tail}}\left(\omega\right)=\frac{ie^{2}}{\hbar^{2}\mathcal{V}\left(\omega+i0^{+}\right)}\sum_{\boldsymbol{k}n}f_{\boldsymbol{k}n}\\
\sum_{\nu_{1}\nu_{2}}\Omega_{\boldsymbol{k}\nu_{1}n}^{*}\left[\partial_{k_{\parallel}}^{2}\tilde{T}_{\boldsymbol{k}}^{\nu_{1}\nu_{2}}\right]\Omega_{\boldsymbol{k}\nu_{2}n},
\end{multline}
where $k_{\parallel}=\boldsymbol{u}\cdot\boldsymbol{k}$. Here we
consider a fully gapped system (e.g., a semiconductor) at zero temperature,
so that $f_{\boldsymbol{k}n}$ doesn't depend on $\boldsymbol{k}$,
but it is just a function of $n$, and can be only 1 or 0. As such,
$\partial_{k_{\parallel}}f_{\boldsymbol{k}n}=0$, and one can use
the periodicity of the FBZ to perform an integration by part and write
$\sigma_{2}^{\mathrm{eq},\mathrm{tail}}\left(\omega\right)$ as,

\begin{align}
 & \sigma_{2}^{\mathrm{eq},\mathrm{tail}}\left(\omega\right)=-\frac{ie^{2}}{\hbar^{2}\mathcal{V}\left(\omega+i0^{+}\right)}\sum_{\boldsymbol{k}n}f_{\boldsymbol{k}n}\nonumber \\
 & \left\{ \left[\partial_{k_{\parallel}}\Omega_{\boldsymbol{k}}^{\dagger}\right]\Omega_{\boldsymbol{k}}\eta_{\boldsymbol{k}}+\eta_{\boldsymbol{k}}\Omega_{\boldsymbol{k}}^{\dagger}\left[\partial_{k_{\parallel}}\Omega_{\boldsymbol{k}}\right]\right\} _{nn}
\end{align}
where, for the sake of simplicity, a compact matrix notation has been
used, $\left\{ M\right\} _{nn^{\prime}}$ is the $nn^{\prime}$ element
of matrix $M$, and $\eta_{\boldsymbol{k}}=\boldsymbol{u}\cdot\boldsymbol{\eta}_{\boldsymbol{k}}=\Omega_{\boldsymbol{k}}^{\dagger}\left[\partial_{k_{\parallel}}\tilde{T}_{\boldsymbol{k}}\right]\Omega_{\boldsymbol{k}}$.
Then, we exploit the following relation \citep{yates2007spectral},
\begin{equation}
\left\{ \Omega_{\boldsymbol{k}}^{\dagger}\left[\partial_{k_{\parallel}}\Omega_{\boldsymbol{k}}\right]\right\} _{nn^{\prime}}=\begin{cases}
-\frac{\eta_{\boldsymbol{k},nn^{\prime}}}{\hbar\omega_{\boldsymbol{k},nn^{\prime}}} & n\neq n^{\prime},\\
0 & n=n^{\prime},
\end{cases}\label{eq:U-dag-U}
\end{equation}
to obtain

\begin{align}
 & \sigma_{2}^{\mathrm{eq},\mathrm{tail}}\left(\omega\right)=-\frac{ie^{2}}{\hbar^{3}\mathcal{V}\left(\omega+i0^{+}\right)}\sum_{\boldsymbol{k}n}f_{\boldsymbol{k}n}\nonumber \\
 & \sum_{n^{\prime}\neq n}\left(\frac{\eta_{\boldsymbol{k},nn^{\prime}}\eta_{\boldsymbol{k},n^{\prime}n}}{\omega_{\boldsymbol{k},nn^{\prime}}}-\frac{\eta_{\boldsymbol{k},nn^{\prime}}\eta_{\boldsymbol{k},n^{\prime}n}}{\omega_{\boldsymbol{k},n^{\prime}n}}\right),
\end{align}
which after some straightforward calculations can be rewritten as
\begin{align}
 & \sigma_{2}^{\mathrm{eq},\mathrm{tail}}\left(\omega\right)=\frac{ie^{2}}{\hbar^{3}\mathcal{V}\left(\omega+i0^{+}\right)}\sum_{\boldsymbol{k}nn^{\prime}}\nonumber \\
 & \frac{\eta_{\boldsymbol{k},nn^{\prime}}\eta_{\boldsymbol{k},n^{\prime}n}\left(f_{\boldsymbol{k}n^{\prime}}-f_{\boldsymbol{k}n}\right)}{\omega_{\boldsymbol{k},nn^{\prime}}}.\label{eq:sigma-eq-diag-0T-tail-2}
\end{align}
Comparing Eqs.~\ref{eq:sigma-eq-tail-1} and \ref{eq:sigma-eq-diag-0T-tail-2},
one verifies that
\begin{equation}
\sigma_{1}^{\mathrm{eq},\mathrm{tail}}\left(\omega\right)+\sigma_{2}^{\mathrm{eq},\mathrm{tail}}\left(\omega\right)=0.
\end{equation}
 A crucial point in this proof is that the distribution function,$f_{\boldsymbol{k}n}$,
is independent of $\boldsymbol{k}$. This is clearly the case for
fully gapped systems at zero temperature, where indeed we have no
DC conductivity and the tails should cancel each other which results
in a vanishing Drude weight \citep{schuler2021gauge}. On the other
hand, if the temperature is not zero or the system is metallic (non-gapped),
this condition breaks down and we can have DC conductivity as the
tail terms don't cancel each other.

\begin{figure*}[t]
\centering{}\includegraphics[width=18cm]{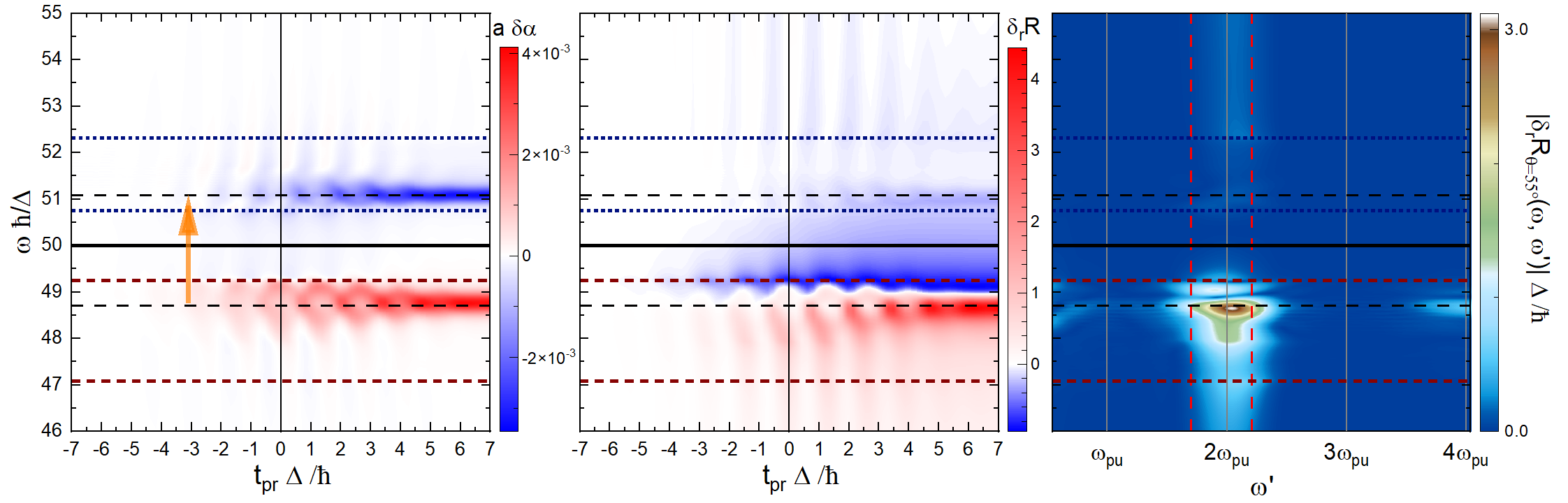}\caption{Transient optical behavior of the pumped system, computed using the
quasi-static approximation. All parameters are similar to the ones
of Fig.~\ref{fig:dR55_D=00003D0}. (left) The transient differential
absorption coefficient, (middle) the transient relative differential
reflectivity at the incident angle of $\theta=55{^\circ}$, and (left)
the amplitude of the Fourier transformation of the transient reflectivity,
$\left|\delta R_{\theta=55{^\circ}}\left(\omega,\omega^{\prime}\right)\right|$.\protect\label{fig:dR55_D=00003D0_quasi-static}}
\end{figure*}

\section{Quasi-static approximation\protect\label{sec:Quasi-static-approximation}}

One approximation is to assume that the probe frequency is much higher
than the one of the pump pulse, so that one can consider the system
to be quasi static in the vicinity of the probing time $t_{\mathrm{pr}}$.
For such an approximation, one computes the dynamics up to $t_{\mathrm{pr}}$
according to Eq.~\ref{eq:P-dyn-gen} to obtain $P_{\boldsymbol{k}}\left(t_{\mathrm{pr}}\right)$,
but, for those times larger than $t_{\mathrm{pr}}$, one considers
the Hamiltonian as time independent. Diagonalizing $H_{\boldsymbol{k}}^{\prime}\left(t_{\mathrm{pr}}\right)$
using a unitary matrix $U_{\boldsymbol{k}}\left(t_{\mathrm{pr}}\right)$,
one can simply show that the quasi-static projection coefficients
in this diagonal basis, $P_{\boldsymbol{k}ln}^{\mathrm{qs}}\left(t\right)$,
are given by 
\begin{align}
P_{\boldsymbol{k}ln}^{\mathrm{qs}}\left(t\right)=e^{-i\left(t-t_{\mathrm{pr}}\right)\omega_{\boldsymbol{k}l}\left(t_{\mathrm{pr}}\right)}P_{\boldsymbol{k}ln}^{\mathrm{qs}}\left(t_{\mathrm{pr}}\right)\qquad & t\geq t_{\mathrm{pr}}\label{eq:P-qs}
\end{align}
where $\hbar\omega_{\boldsymbol{k}l}\left(t_{\mathrm{pr}}\right)$
is the $l$-th eigenvalue of $H_{\boldsymbol{k}}^{\prime}\left(t_{\mathrm{pr}}\right)$,
and 
\begin{equation}
P_{\boldsymbol{k}ln}^{\mathrm{qs}}\left(t_{\mathrm{pr}}\right)=\sum_{n^{\prime}}U_{\boldsymbol{k}n^{\prime}l}^{\star}\left(t_{\mathrm{pr}}\right)P_{\boldsymbol{k}n^{\prime}n}\left(t_{\mathrm{pr}}\right).
\end{equation}
Moreover, all of the coupling matrices $\boldsymbol{J}$, $\boldsymbol{V}$
, $\boldsymbol{D}$ and $\delta\boldsymbol{J}/\delta\boldsymbol{A}$
should be evaluated at the time $t_{\mathrm{pr}}$ and transformed
using $U_{\boldsymbol{k}}\left(t_{\mathrm{pr}}\right)$. Performing
some algebra, one obtains the two parts of quasi-static conductivity
as,

\begin{align}
 & \boldsymbol{\sigma}_{1}^{\mathrm{qs}}\left(\omega,t_{\mathrm{pr}}\right)=\frac{ie^{2}}{\hbar\mathcal{V}}\sum_{\boldsymbol{k}}\sum_{l_{1}l_{2}l_{3}}\nonumber \\
 & \left\{ -\frac{\boldsymbol{J}_{\boldsymbol{k},l_{2}l_{1}}\left(t_{\mathrm{pr}}\right)\boldsymbol{V}_{\boldsymbol{k},l_{1}l_{3}}\left(t_{\mathrm{pr}}\right)N_{\boldsymbol{k},l_{2}l_{3}}\left(t_{\mathrm{pr}}\right)}{\omega_{\boldsymbol{k}l_{1}l_{3}}\left(t_{\mathrm{pr}}\right)\left[\omega_{\boldsymbol{k}l_{2}l_{3}}\left(t_{\mathrm{pr}}\right)+\omega+i0^{+}\right]}\right.\nonumber \\
 & \left.+\frac{\boldsymbol{J}_{\boldsymbol{k},l_{2}l_{1}}\left(t_{\mathrm{pr}}\right)\boldsymbol{V}_{\boldsymbol{k},l_{1}l_{3}}\left(t_{\mathrm{pr}}\right)N_{\boldsymbol{k},l_{2}l_{3}}\left(t_{\mathrm{pr}}\right)}{\omega_{\boldsymbol{k}l_{1}l_{3}}\left(t_{\mathrm{pr}}\right)\left[\omega_{\boldsymbol{k}l_{2}l_{1}}\left(t_{\mathrm{pr}}\right)+\omega+i0^{+}\right]}\right.\nonumber \\
 & \left.+\frac{\boldsymbol{J}_{\boldsymbol{k},l_{3}l_{2}}\left(t_{\mathrm{pr}}\right)\boldsymbol{V}_{\boldsymbol{k},l_{1}l_{3}}\left(t_{\mathrm{pr}}\right)N_{\boldsymbol{k},l_{1}l_{2}}\left(t_{\mathrm{pr}}\right)}{\omega_{\boldsymbol{k}l_{1}l_{3}}\left(t_{\mathrm{pr}}\right)\left[\omega_{\boldsymbol{k}l_{1}l_{2}}\left(t_{\mathrm{pr}}\right)+\omega+i0^{+}\right]}\right.\nonumber \\
 & \left.-\frac{\boldsymbol{J}_{\boldsymbol{k},l_{3}l_{2}}\left(t_{\mathrm{pr}}\right)\boldsymbol{V}_{\boldsymbol{k},l_{1}l_{3}}\left(t_{\mathrm{pr}}\right)N_{\boldsymbol{k},l_{1}l_{2}}\left(t_{\mathrm{pr}}\right)}{\omega_{\boldsymbol{k}l_{1}l_{3}}\left(t_{\mathrm{pr}}\right)\left[\omega_{\boldsymbol{k}l_{3}l_{2}}\left(t_{\mathrm{pr}}\right)+\omega+i0^{+}\right]}\right.\nonumber \\
 & \left.+\frac{i\boldsymbol{J}_{\boldsymbol{k},l_{2}l_{1}}\left(t_{\mathrm{pr}}\right)\boldsymbol{D}_{\boldsymbol{k},l_{1}l}\left(t_{\mathrm{pr}}\right)N_{\boldsymbol{k},l_{2}l_{3}}\left(t_{\mathrm{pr}}\right)}{\omega_{\boldsymbol{k}l_{2}l_{1}}\left(t_{\mathrm{pr}}\right)+\omega+i0^{+}}\right.\nonumber \\
 & \left.-\frac{i\boldsymbol{J}_{\boldsymbol{k},l_{3}l_{2}}\left(t_{\mathrm{pr}}\right)\boldsymbol{D}_{\boldsymbol{k},l_{1}l_{3}}\left(t_{\mathrm{pr}}\right)N_{\boldsymbol{k},l_{1}l_{2}}\left(t_{\mathrm{pr}}\right)}{\omega_{\boldsymbol{k}l_{3}l_{2}}\left(t_{\mathrm{pr}}\right)+\omega+i0^{+}}\right\} ,\label{eq:sigma-qs-1}
\end{align}
and
\begin{equation}
\boldsymbol{\sigma}_{2}^{\mathrm{qs}}\left(\omega,t_{\mathrm{pr}}\right)=\frac{e}{\mathcal{V}}\sum_{\boldsymbol{k}}\sum_{l_{1}l_{2}}\frac{iN_{\boldsymbol{k},l_{1}l_{2}}\left(t_{\mathrm{pr}}\right)\frac{\delta\boldsymbol{J}_{\boldsymbol{k},l_{1}l_{2}}\left(t_{\mathrm{pr}}\right)}{\delta\boldsymbol{A}}}{\omega_{\boldsymbol{k}l_{1}l_{2}}\left(t_{\mathrm{pr}}\right)+\omega+i0^{+}},\label{eq:sigma-qs-2}
\end{equation}
where $\omega_{\boldsymbol{k}ll^{\prime}}\left(t_{\mathrm{pr}}\right)=\omega_{\boldsymbol{k}l}\left(t_{\mathrm{pr}}\right)-\omega_{\boldsymbol{k}l^{\prime}}\left(t_{\mathrm{pr}}\right)$
is the difference between eigenenergies of the time-dependent single-particle
Hamiltonian at time $t_{\mathrm{pr}}$, $N_{\boldsymbol{k}ll^{\prime}}\left(t_{\mathrm{pr}}\right)=\sum_{n}P_{\boldsymbol{k}ln}^{\mathrm{qs}\star}\left(t_{\mathrm{pr}}\right)P_{\boldsymbol{k}l^{\prime}n}^{\mathrm{qs}}\left(t_{\mathrm{pr}}\right)f_{\boldsymbol{k}n}$
is the dynamical population matrix of the corresponding eigenbasis,
and in the derivation we have used the following property of the projection
coefficients: $\sum_{n}P_{\boldsymbol{k}ln}^{\mathrm{qs}\star}\left(t_{\mathrm{pr}}\right)P_{\boldsymbol{k}l^{\prime}n}^{\mathrm{qs}}\left(t_{\mathrm{pr}}\right)=\delta_{ll^{\prime}}$.

For the pumped toy-model with the parameters of Fig.~\ref{fig:dR55_D=00003D0}
in the main text, we calculate the transient optical behavior, using
quasi-static approximation and the results are shown in Fig.~\ref{fig:dR55_D=00003D0_quasi-static}.
Clearly, even though there are some similarities between the results
presented in Figs.~\ref{fig:dR55_D=00003D0_quasi-static} and \ref{fig:dR55_D=00003D0},
many of the details of the full calculations are not captured in the
quasi-static approximation.

\begin{figure}[t]
\centering{}\includegraphics[width=6cm]{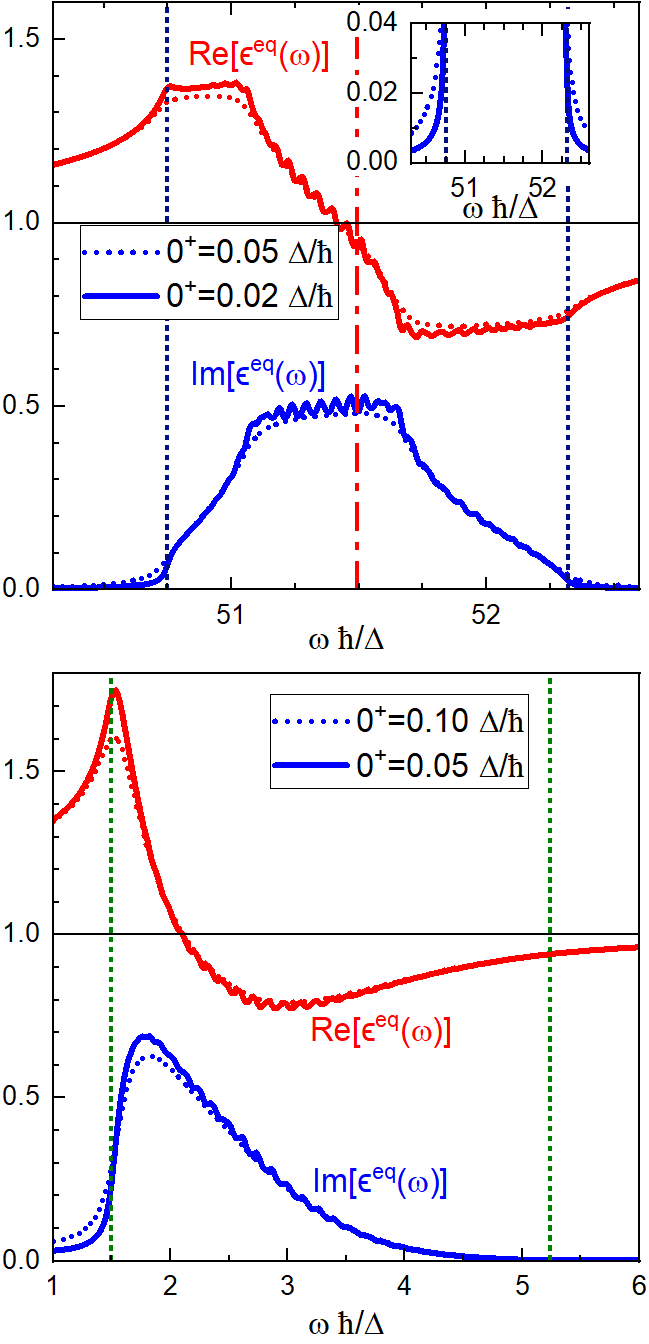}\caption{(top) The equilibrium dielectric function in the XUV regime. The solid
curves correspond to $0^{+}=0.02\Delta/\hbar$, while the dotted curves
correspond to $0^{+}=0.05\Delta/\hbar$, which is used in the XUV
calculations in the main text. The inset magnifies the low values
of the y axis and shows how the smaller $0^{+}$ makes $\Im\left[\epsilon^{\mathrm{eq}}\left(\omega\right)\right]$
vanish more rapidly outside of the energy range of the CB. (bottom)
The equilibrium dielectric function in the IR-V regime. The solid
curves correspond to $0^{+}=0.05\Delta/\hbar$, while the dotted curves
correspond to $0^{+}=0.1\Delta/\hbar$, as used in Fig.~\ref{fig:IRV}.
\protect\label{fig:zp_stud}}
\end{figure}

\section{A brief study on the damping factor ($0^{+}$)\protect\label{sec:zp_study}}

As described in the main text (see Sec.~\ref{subsec:Finite-k-grid}),
$0^{+}$ should be chosen in such a way to make individual \textbf{k}
points on the grid indistinguishable. As such, for the case of the
XUV probe, we chose it to be $0.05\Delta/\hbar$. In Fig.~\ref{fig:zp_stud}
top panel, we plot the equilibrium dielectric function in the XUV
regime, with $0^{+}=0.02\Delta/\hbar$ and compare it with the case
of $0^{+}=0.05\Delta/\hbar$ , which is used in all of the XUV calculations
in the main text. Clearly, the small value of $0^{+}=0.02\Delta/\hbar$
introduces fluctuations which are the distinguishable effects of individual
\textbf{k} points on the grid. The inset magnifies the low values
of the y axis and shows how the smaller $0^{+}$ makes $\Im\left[\epsilon^{\mathrm{eq}}\left(\omega\right)\right]$
vanish more rapidly outside of the energy range of the CB. This confirms
our explanation that the broadening of $\Im\left[\epsilon^{\mathrm{eq}}\left(\omega\right)\right]$
outside of the energy range of the CB is because of the finite value
of $0^{+}$.

Going to the IR-V regime, we work with lower frequencies which makes
the results more sensitive to the energy differences between adjacent
\textbf{k} points on the grid and calls for a larger damping factor.
In Fig.~\ref{fig:zp_stud} bottom panel, we plot the equilibrium
dielectric function in the IR-V regime, with $0^{+}=0.05\Delta/\hbar$
and compare it with the case of $0^{+}=0.1\Delta/\hbar$ , which is
used in the IR-V calculations in the main text. Clearly, even though
the value $0^{+}=0.05\Delta/\hbar$ works well for the XUV regime,
for the IR-V regime it is not adequate. On the other hand, using the
more reliable value of $0^{+}=0.1\Delta/\hbar$ has the problem that
for the low frequencies outside of the gap-energy range, $\Im\left[\epsilon^{\mathrm{eq}}\left(\omega\right)\right]$
doesn't vanish rapidly enough, and there is no way out of such limitations.

\begin{figure}[t]
\centering{}\includegraphics[width=6cm]{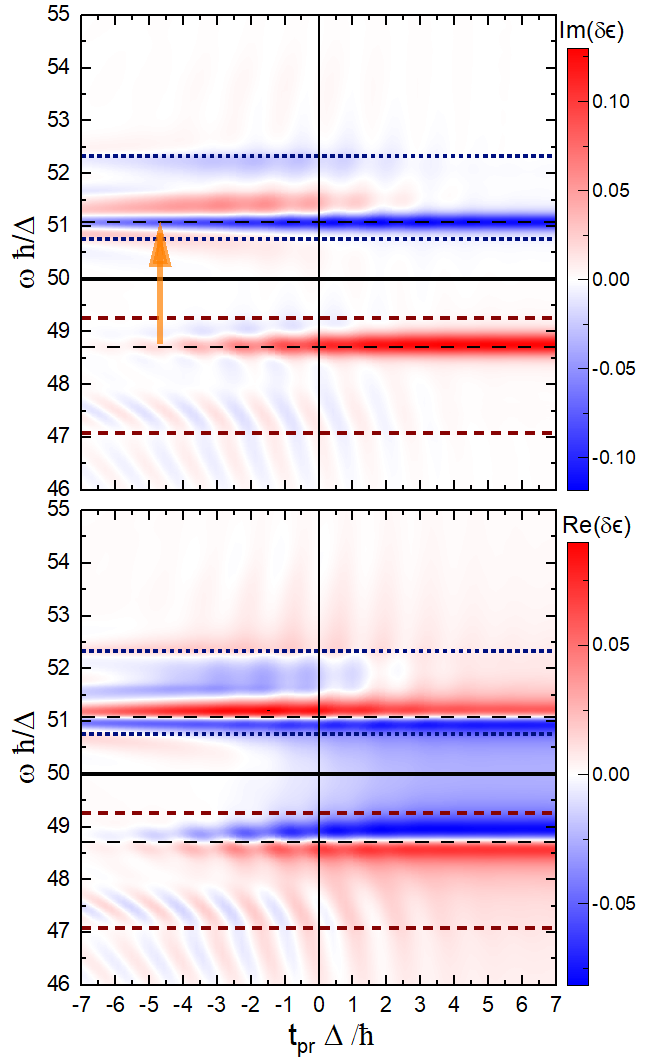}\caption{Transient behavior of the differential (left) imaginary part of the
dielectric function, $\delta\Im\epsilon\left(\omega,t_{\mathrm{pr}}\right)$,
and (right) its real part, $\delta\Re\epsilon\left(\omega,t_{\mathrm{pr}}\right)$,
vs the probe time and photon energy, with the same parameters as the
ones of Fig.~\ref{fig:dR55_D=00003D0}. \protect\label{fig:d_eps_D=00003D0}}
\end{figure}

\section{Transient differential dielectric function\protect\label{sec:differential-epsilon}}

In this appendix, we show the transient differential imaginary and
real parts of the dielectric function, i.e.,

\begin{equation}
\delta\Im\epsilon\left(\omega,t_{\mathrm{pr}}\right)=\Im\left[\epsilon\left(\omega,t_{\mathrm{pr}}\right)-\epsilon^{\mathrm{eq}}\left(\omega\right)\right],
\end{equation}
and

\begin{equation}
\delta\Re\epsilon\left(\omega,t_{\mathrm{pr}}\right)=\Re\left[\epsilon\left(\omega,t_{\mathrm{pr}}\right)-\epsilon^{\mathrm{eq}}\left(\omega\right)\right],
\end{equation}
respectively, where $\epsilon^{\mathrm{eq}}\left(\omega\right)=\epsilon\left(\omega,t_{\mathrm{pr}}\rightarrow t_{\mathrm{ini}}\right)$
is the equilibrium dielectric function.

Fig.~\ref{fig:d_eps_D=00003D0} shows the transient behavior of the
differential imaginary and real parts of the dielectric function,
with the same parameters as the ones of Fig.~\ref{fig:dR55_D=00003D0}.
Comparison between Figs.~\ref{fig:dR55_D=00003D0} and \ref{fig:d_eps_D=00003D0},
one can see that the behavior of the absorption coefficient can be
qualitatively very well understood by studying just $\Im\epsilon\left(\omega,t_{\mathrm{pr}}\right)$,
as it is also clear from the Eq.~\ref{eq:alpha}. Moreover, the transient
reflectivity behavior can also be qualitatively understood to some
extent from the transient real part of the dielectric function, $\Re\epsilon\left(\omega,t_{\mathrm{pr}}\right)$,
as the latter is the main ingredient of the refractive index and hence
plays a major role in determining the former.

\begin{figure*}[t]
\centering{}\includegraphics[width=18cm]{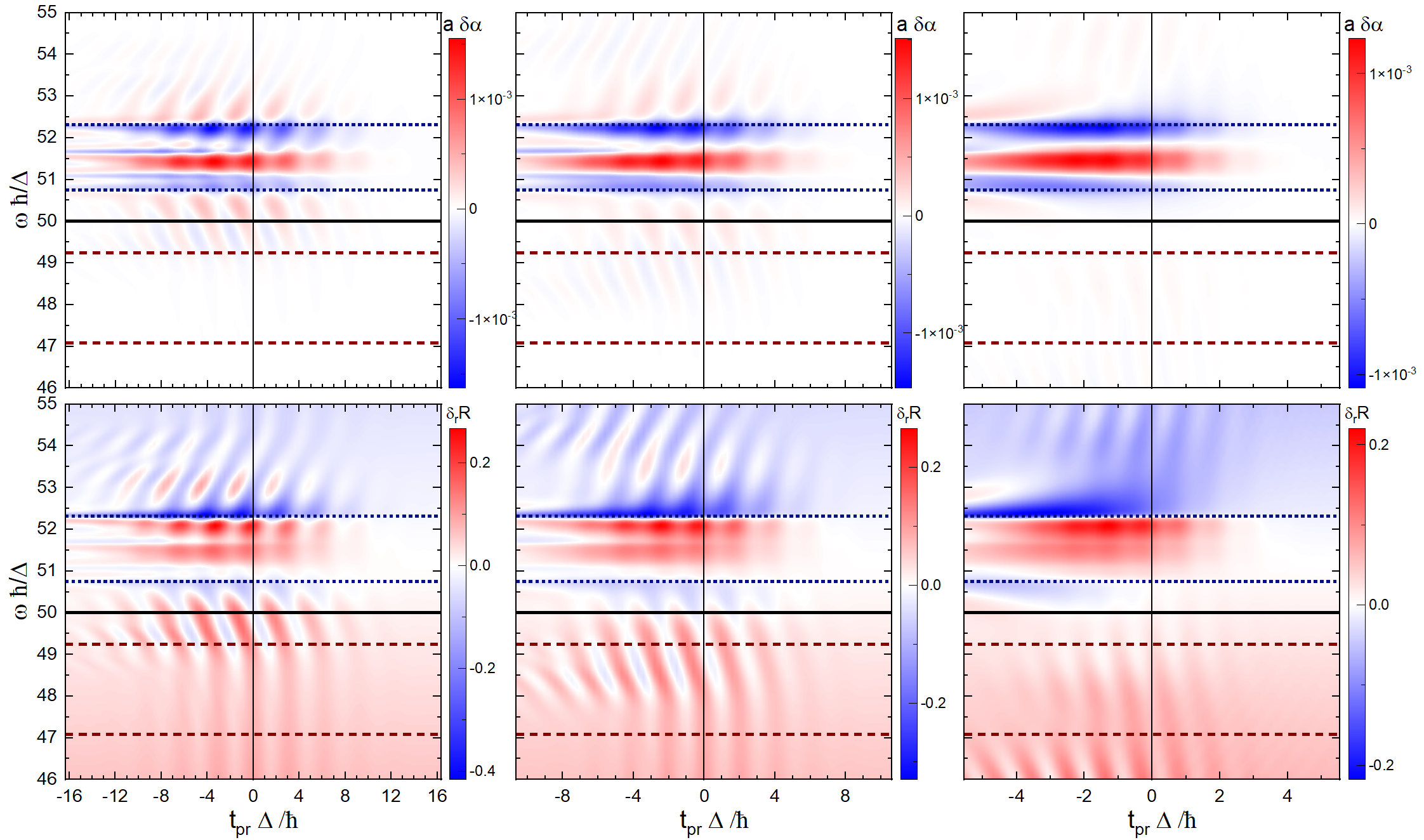}\caption{Only intra-band transitions in the dynamics. (top row) Transient differential
absorption coefficients and (bottom row) transient relative differential
reflectivities for three different pump-pulse frequencies: (left column)
$\hbar\omega_{\mathrm{pu}}/\Delta=1$, (middle column)$\hbar\omega_{\mathrm{pu}}/\Delta=1.5$
and (right column) $\hbar\omega_{\mathrm{pu}}/\Delta=3$, and the
rest of the parameter are similar to the corresponding panels of Fig.~\ref{fig:dR55_D=00003D0_wps}.
\protect\label{fig:dR55_D=00003D0_wps-intra}}
\end{figure*}

\section{Density of states and its generalizations\protect\label{sec:Density-of-states}}

We define the density of states, $g\left(\varepsilon\right)$, at
each energy $\varepsilon$ as the number of states within the infinitesimal
range $d\varepsilon$ around $\varepsilon$ divided by $d\varepsilon$,
in the volume of the unit cell. Even though mathematically, $g\left(\varepsilon\right)$
is defined using a Dirac-delta function, in numerical calculations
it is obtained as, 
\begin{equation}
g\left(\varepsilon\right)=\frac{1}{N_{\mathrm{grid}}}\sum_{\boldsymbol{k}\in\mathrm{grid}}\sum_{n}L\left(\varepsilon-\varepsilon_{\mathbf{k},n}\right),
\end{equation}
where $L\left(\varepsilon\right)=\frac{\hbar0^{+}}{\pi\left(\hbar^{2}0^{+2}+\varepsilon^{2}\right)}$
is the Lorentzian function. This concept can be generalized to any
momentum and band dependent property, $O_{\mathbf{k},n}$, such as
diagonal velocity and inverse-mass, etc., with the density given by:

\begin{equation}
g_{O}\left(\varepsilon\right)=\frac{1}{N_{\mathrm{grid}}}\sum_{\boldsymbol{k}\in\mathrm{grid}}\sum_{n}O_{\mathbf{k},n}L\left(\varepsilon-\varepsilon_{\mathbf{k},n}\right).
\end{equation}

Another generalization is to compute the $n_{\mathrm{ph}}$-photon
density of resonant states (DORS). Roughly speaking, the $n_{\mathrm{ph}}$
transitions come from the $n_{\mathrm{ph}}$-th power of the pumping
field. We define the function $w_{n_{\mathrm{ph}}}\left(\varepsilon_{\mathrm{gap}}\right)$
which gives the strength of the $n_{\mathrm{ph}}$ resonance for each
given gap-energy $\varepsilon_{\mathrm{gap}}$, and is obtained from
the leading term of the square of the amplitude of the $\varepsilon_{\mathrm{gap}}/\hbar$
component in the spectrum of the $n_{\mathrm{ph}}$-th power of the
pump pulse, centered at $n_{\mathrm{ph}}\omega_{\mathrm{pu}}$. With
our pump, Eq.~\ref{eq:A_t}, it is given by \citep{eskandari2023dynamical}

\begin{align}
w_{n_{\mathrm{ph}}}\left(\varepsilon_{\mathrm{gap}}\right)=e^{-\frac{\tau_{\mathrm{pu}}^{2}}{8n_{\mathrm{ph}}\ln2}\left(\frac{\varepsilon_{\mathrm{gap}}}{\hbar}-n_{\mathrm{ph}}\omega_{\mathrm{pu}}\right)^{2}}.\label{eq:strenght_res_l}
\end{align}
This strength function is normalized so that at the exact resonance,
$\varepsilon_{\mathrm{gap}}=\hbar n_{\mathrm{ph}}\omega_{\mathrm{pu}}$,
it is equal to 1. Note that, instead of the amplitude, its square
is used, as in a pure Rabi-oscillation with low Rabi frequency, the
excitation population is proportional to the square of the pump amplitude
\citep{eskandari2023dynamical}. Nevertheless, one should be aware
of the fact that different frequency components of the pump pulse
do not act independently, and hence our resonance assignment should
be considered as an approximation. Usually, the gaps between the filled
and empty bands are of interest. At zero temperature, the $n_{\mathrm{ph}}$-photon
DORS is given by

\begin{align}
g_{n_{\mathrm{ph}}}\left(\varepsilon\right) & =\frac{1}{N_{\mathrm{grid}}}\sum_{\boldsymbol{k}\in\mathrm{grid}}\sum_{n_{C}}L\left(\varepsilon-\varepsilon_{\mathbf{k},n_{C}}\right)\nonumber \\
 & \sum_{n_{V}}w_{n_{\mathrm{ph}}}\left(\varepsilon_{\mathbf{k},n_{C}}-\varepsilon_{\mathbf{k},n_{V}}\right),\qquad\varepsilon>\varepsilon_{F}
\end{align}
\begin{align}
g_{n_{\mathrm{ph}}}\left(\varepsilon\right) & =\frac{1}{N_{\mathrm{grid}}}\sum_{\boldsymbol{k}\in\mathrm{grid}}\sum_{n_{V}}L\left(\varepsilon-\varepsilon_{\mathbf{k},n_{V}}\right)\nonumber \\
 & \sum_{n_{C}}w_{n_{\mathrm{ph}}}\left(\varepsilon_{\mathbf{k},n_{C}}-\varepsilon_{\mathbf{k},n_{V}}\right),\qquad\varepsilon<\varepsilon_{F}
\end{align}
where $n_{C}$ ($n_{V}$) runs over the CBs (VBs) and $\varepsilon_{F}$
is the Fermi energy. For finite temperatures, we consider the gaps
between the filled and empty electronic states, and consequently,
generalize $g_{n_{\mathrm{ph}}}\left(\varepsilon\right)$ by obtaining
it as,

\begin{align}
g_{n_{\mathrm{ph}}}\left(\varepsilon\right) & =\frac{1}{N_{\mathrm{grid}}}\sum_{\boldsymbol{k}\in\mathrm{grid}}\sum_{nn^{\prime}}\nonumber \\
 & \left[f_{\mathbf{k},n}\left(1-f_{\mathbf{k},n^{\prime}}\right)+f_{\mathbf{k},n^{\prime}}\left(1-f_{\mathbf{k},n}\right)\right]\nonumber \\
 & L\left(\varepsilon-\varepsilon_{\mathbf{k},n}\right)w_{n_{\mathrm{ph}}}\left(\left|\varepsilon_{\mathbf{k},n}-\varepsilon_{\mathbf{k},n^{\prime}}\right|\right).
\end{align}

\section{Intra-band motion with different pump-pulse frequencies\protect\label{sec:Intra-band}}

In Fig.~\ref{fig:dR55_D=00003D0_wps-intra}, we show the transient
differential absorption coefficient and relative differential reflectivity
with considering only intra-band transitions upon pumping, for three
different pump frequencies, $\hbar\omega_{\mathrm{pu}}/\Delta=$1,
1.5, 3, similar to Fig.~\ref{fig:dR55_D=00003D0_wps}. Clearly, the
color maps are qualitatively very similar, i.e., the probe photon
energies at which positive or negative (red or blue, respectively)
features appear, are almost independent of the frequency of the pump
pulse, even though the details of the oscillations vary by varying
it.

\bibliographystyle{apsrev4-2}
\bibliography{biblio}

\end{document}